\documentclass[a4paper,onecolumn,11pt,unpublished]{quantumarticle}
\pdfoutput=1
\usepackage[utf8]{inputenc}
\usepackage[english]{babel}
\usepackage[T1]{fontenc}
\usepackage{orcidlink}
\usepackage{amsmath,amssymb,amsthm,mathtools,microtype}
\usepackage{booktabs,array,enumitem,placeins}
\usepackage[numbers,sort&compress]{natbib}
\usepackage{hyperref}
\AtBeginDocument{\normalfont}

\hypersetup{
  allcolors=quantumviolet,
  pdftitle={Query-Optimal Quantum Minimum Finding from Numerical Estimators},
}
\setlist[enumerate]{itemsep=3pt,topsep=5pt}
\numberwithin{equation}{section}
\newtheorem{theorem}{Theorem}[section]
\newtheorem{lemma}[theorem]{Lemma}
\newtheorem{proposition}[theorem]{Proposition}
\newtheorem{corollary}[theorem]{Corollary}
\theoremstyle{definition}
\newtheorem{definition}[theorem]{Definition}
\newtheorem{algorithm}{Algorithm}
\theoremstyle{remark}
\newtheorem{remark}[theorem]{Remark}
\newcommand{\ket}[1]{\lvert #1\rangle}
\newcommand{\bra}[1]{\langle #1\rvert}
\newcommand{\norm}[1]{\lVert #1\rVert}
\newcommand{\eps}{\varepsilon}
\newcommand{\Tr}{\operatorname{Tr}}
\newcommand{\polylog}{\operatorname{polylog}}
\newcommand{\Val}{\operatorname{Val}}

\newcommand{\one}{\mathbf1}
\newcommand{\braket}[1]{\langle #1 \rangle}

\title{Query-Optimal Quantum Minimum Finding from Numerical Estimators}

\author[1]{Egil Bae\,\orcidlink{0000-0002-4621-259X}}
\email{Egil.Bae@ffi.no}
\affil[1]{Norwegian Defence Research Establishment (FFI), Kjeller, Norway}

\begin{document}
\maketitle
\begin{abstract}

We propose a quantum minimum-finding algorithm for objectives accessed via numerical estimators that avoids the logarithmic query overhead of amplifying the confidence of estimates before the search. A coherent estimator that is $\varepsilon$-accurate with probability at least $2/3$ on each of $N$ labels suffices to find a $2\varepsilon$-optimal label using $O(\sqrt N)$ base-estimator queries at constant confidence. The query bound and the additive-error constant are separately worst-case optimal. More generally, suppose a preparation assigns probability mass at least a known $p$ to objective values at most $a$. Without knowing $a$ or the prepared distribution, our algorithm returns a candidate and an estimate $v$ such that, jointly with constant probability, $v$ is $\varepsilon$-accurate for the candidate and $v\le a+\varepsilon$. The algorithm uses $O(p^{-1/2})$ queries to each of the preparation and estimator oracles in the worst case, and the selected candidate can be retained as a quantum state. For preparation width $s_P$ and estimator workspace width $w$, it requires $O(s_P+w+\log\log(4/p))$ qubits, independently of the requested confidence. Applications give worst-case optimal $\Theta(\sqrt N/\eta)$ queries for selecting an $\eta$-optimal bounded expectation, and guided ground-energy estimation using $O(\omega^{-1})$ preparations and $O((\omega\eta)^{-1})$ elementary controlled-evolution queries at constant confidence, where $\omega$ is a lower bound on the guiding state's ground-space amplitude overlap.

\end{abstract}
\section{Introduction}\label{sec:intro}

Quantum minimum finding extends the quadratic advantage of quantum search~\cite{Grover,BBBV,BBHT} to finding the smallest of $N$ arbitrary values using $O(\sqrt N)$ evaluations~\cite{DH}. In numerical applications, the objective value may not be available directly, and each evaluation may itself involve an estimation procedure. For example, coherent simulation can be used to estimate an expected loss, while phase estimation can be used to estimate the energy of a quantum state. In this work, we consider estimators that, for every candidate, return a value within additive error $\eps$ of the objective with probability at least $2/3$.

To avoid biasing the search towards poor candidates with low estimated values, a standard approach applies the estimator several times and takes the median of the outputs. Implemented coherently, this procedure can reduce the estimator's failure probability to $O(1/N)$ using $O(\log N)$ repetitions. For bounded-error Boolean search, H{\o}yer, Mosca, and de Wolf~\cite{HMW} showed that this logarithmic overhead can be avoided by interleaving amplitude amplification with error reduction. In case of numerical estimation, however, the probability of obtaining an accurate estimate may be distributed among many accurate output values, such that no individual value has probability greater than $1/2$. Moreover, accurate estimates may give different orderings of candidates with similar objective values, or place the same candidate on opposite sides of a threshold. The probability that a candidate passes a threshold comparison depends on how often its estimates fall below the threshold. Lowering the threshold can reduce the acceptance probability by different amounts for different candidates and consequently change their relative probabilities of being selected.

Generalized minimum finding considers the problem of selecting a candidate with a low encoded score from an arbitrary prepared distribution, without requiring knowledge of the qualifying threshold~\cite{AGGW}. The procedure requires $O(p^{-1/2})$ queries to the score-preparation unitary and its inverse to succeed with constant confidence, where $p$ is a known lower bound on the prepared probability mass of scores at most this threshold. When objective values are available only through numerical estimators, a standard approach is to first amplify the confidence of the estimator. More specifically, its failure probability is reduced to a sufficiently small constant multiple of $p$, resulting in a total cost of $O(p^{-1/2}\log(2/p))$ base-estimator queries. Chen and de Wolf~\cite{CDW} use this approach to derive their approximate-unitary minimum-finding theorem from the generalized procedure, with a uniform preparation over $N$ labels and $p=1/N$. The same reduction is also used in the $k=1$ special case of Gao, Ji, and Wang~\cite{GJW}. With confidence amplification, both these approaches therefore use $O(\sqrt N\log(2N))$ base-estimator queries and give an additive selection error of $2\eps$. The factor $2$ in this error bound cannot in general be improved, as we prove in Proposition~\ref{prop:accuracy-lower}.

We propose a quantum minimum-finding algorithm for general objectives accessed through numerical estimators that avoids the logarithmic overhead and attains the optimal additive selection error of $2\eps$. For arbitrary prepared distributions, the algorithm achieves simultaneous $O(p^{-1/2})$ bounds on preparation and base-estimator queries at constant confidence. It returns a candidate together with an accurate estimate of its objective value. The selected candidate can be returned as a classical label or as a quantum state.

\subsection{Main results}

Consider first the special case of $N$ classical labels and a coherent estimator that is $\eps$-accurate with probability at least $2/3$ on every label. The objective values can be arbitrarily close and there can be arbitrarily many ties. We propose an algorithm that with constant confidence returns a label $s$ such that \[ f(s)\le\min_i f(i)+2\eps. \] The algorithm requires $O(\sqrt N)$ queries to the base estimator and its inverse. The query scaling and the additive-error constant are individually worst-case optimal for $0<\eps<1/2$.

In the general setting, assume that the estimator has the full unitary form $A=\sum_i\Pi_i\otimes A_i$, where $\Pi_i$ are projectors onto objective sectors with corresponding objective values $f_i$. Within each sector, all input states give the same distribution of estimates, including states entangled with other registers. This form of $A$ is stronger than preserving sector probabilities alone, as explained in Section~\ref{sec:model}.

Let $P$ be a state preparation unitary and assume access to $P$, $A$, and their inverses. Write $\mu\{f\le a\}$ for the probability that a measurement of the prepared state would yield an objective value at most $a$. Assume that $p\in(0,1]$ is known, and define
\[
 a_p=\min\{a:\mu\{f\le a\}\ge p\}.
\]
The value of $a_p$ is used only in the analysis and is not needed by the algorithm.

Theorem~\ref{thm:main} shows that, with probability at least $1-\delta$, the algorithm returns a candidate together with an estimate $v$ of its objective value such that
\[
 |v-X_{\rm out}|\le\eps,\qquad v\le a_p+\eps,
\]
for any fixed $0<\delta<1/2$. Here $X_{\rm out}$ denotes the objective value that would be obtained if the returned candidate had been measured. Writing $u=\log(2/\delta)$, the worst-case bounds on the preparation query count $Q_P$, estimator query count $Q_A$, and number of qubits $S_Q$ are
\[
 Q_P=O(p^{-1/2}u),\qquad Q_A=O(p^{-1/2}u+u^2),\qquad
 S_Q=O(s_P+w+\log\log(4/p)),
 \]
where $s_P$ and $w$ are the number of qubits used for preparation and for the estimator workspace, respectively. The result applies to both classical labels and unknown energy eigenspaces.

The algorithm is applied to bounded-expectation selection and ground-energy estimation. Combined with amplitude estimation, it gives the worst-case optimal bound of $\Theta(\sqrt N/\eta)$ queries to the underlying coherent sampling oracle for selecting an $\eta$-optimal bounded expectation at constant confidence. For ground-energy estimation, the algorithm uses phase estimation and a guiding state with ground-space amplitude overlap at least $\omega$. An additive-$\eta$ estimate can then be obtained using $O(\omega^{-1}u)$ preparation queries and $O((\omega\eta)^{-1}u)$ elementary controlled-evolution queries. No initial energy estimate is needed, and neither application requires a gap assumption.

\subsection{Proof overview}

The algorithm runs a sequence of trials to improve a current candidate (incumbent). Each trial seeks a new candidate with estimated value $y$ below the estimate $\tau$ of the incumbent. The new candidate is accepted if additional estimates of its value pass validation, otherwise it is rejected. The same estimate $y$ can pass validation more often in one sector than in another. We therefore analyze the probability that a trial proposes a candidate from sector $i$ with estimate $y$.

Let $Y_i$ be a new estimate in sector $i$, and write $q_i^-(y)=\Pr\{Y_i\le y\}$ and $q_i^+(y)=\Pr\{Y_i\ge y\}$. Let $\alpha$ and $\beta$ be constants satisfying $1/3<\alpha<\beta<1/2$. If both the probabilities exceed $\alpha$ for a pair $(i,y)$, then $y$ must differ from the objective value in sector $i$ by at most $\eps$. Such a pair $(i,y)$ is called acceptable. A pair is called good if both probabilities are at least $\beta$. In every sector, an estimate gives a good pair with probability at least $1-2\beta$.

The recursion follows Høyer et al.~\cite{HMW} by interleaving amplitude amplification with error reduction. In our construction, error reduction is performed by two coherent polynomial filters that suppress unacceptable pairs while preserving the weight of good pairs. The filters are constructed using linear-combination block encodings and singular value transformation with real polynomials~\cite{GSLW}. If $\pi_{i,y}$ is the initial probability of $(i,y)$ and $e_\ell(i,y)$ its probability of passing the filters at level $\ell$, the probability that a depth-$j$ trial produces the pair $(i,y)$ without recording a rejection is
\[
c_j(\tau)^2\,\omega_j(i,y)\,\mathbf{1}_{\{y<\tau\}},
\qquad
\omega_j(i,y)=\pi_{i,y}\prod_{\ell=0}^{j-1}e_\ell(i,y).
\]
Since the weights \(\omega_j(i,y)\) are independent of \(\tau\), and the amplification factor \(c_j(\tau)\) is common to all pairs, the relative proposal probabilities of the pairs with estimates below the threshold $\tau$ are preserved when it is lowered. This allows us to analyze the progress of the algorithm using weights that do not depend on the previously accepted estimates.

At depth \(j\), we define the rank \(R_j(\tau)\) as the total \(\omega_j\)-weight of acceptable pairs with \(y<\tau\), and aim to reduce it below a prescribed upper bound, called the rank cap. Rank bounds established at earlier depths, together with the suppression of unacceptable pairs, give a lower bound on the amplification factor. If the current rank exceeds its cap, there is a smallest estimated value \(y_*\) such that the total weight of acceptable pairs with estimates at most \(y_*\) reaches the cap. Accepting a proposal with estimate \(y\le y_*\) lowers the threshold to \(y\), bringing the rank below the cap since only pairs with strictly smaller estimates are counted. The acceptable pairs with estimates at most \(y_*\) have total weight at least the cap, so the proposal formula and validation completeness give a lower bound on the probability of such an update in every trial. Repetition therefore brings the rank below each cap with high probability, even when thresholds depend on previous trials or estimated values have ties.

Since the filters retain a fixed proportion of the initial weight of good pairs from sectors with objective value at most \(a_p\), a threshold \(\tau>a_p+\varepsilon\) gives a final rank of at least a constant times \(p\). Choosing the final rank cap below this amount therefore implies \(\tau\le a_p+\varepsilon\). With high probability, validation also ensures that the stored estimate differs from the objective value by at most \(\varepsilon\), giving a final objective value of at most \(a_p+2\varepsilon\).

The filters require $O(j+1)$ estimator queries at level $j$. Each recursive amplification step applies the preceding circuit twice and its inverse once, followed by a new filter. Consequently, the query cost of a depth-$j$ trial is $O(3^{j-1})$. We choose the maximum depth \(K\) such that \(9^{K-1}=\Theta(1/p)\), giving a maximum trial cost of \(O(3^{K-1})=O(p^{-1/2})\). The number of repetitions increases geometrically toward the cheaper lower depths, but at a rate that keeps the total query cost within the same bound. The query cost of validation is of lower order.

A reversible counter records the level of the first rejection, requiring only \(O(\log(K+1))\) qubits. On branches where no rejection has been recorded, the filter workspace is zero and can be reused. Since \(K=O(\log(2/p))\), this gives a total space bound of \(O(s_P+w+\log\log(4/p))\).\nocite{YY}

\subsection{Related work}

\begin{table}[htbp]
\centering\small
\caption{Query bounds for minimum finding at constant overall confidence, using a numerical estimator that is $\eps$-accurate with probability at least $2/3$ on every label. All methods guarantee an additive selection error of at most $2\eps$.  The first two rows concern an arbitrary preparation with qualifying mass $p$ and count preparation queries. The last three rows concern $N$ labels with known uniform preparation, which is free. }
\label{tab:minimum-comparison}
\begin{tabular}{@{}>{\raggedright\arraybackslash}p{.48\textwidth}>{\raggedright\arraybackslash}p{.19\textwidth}>{\raggedright\arraybackslash}p{.25\textwidth}@{}}
\toprule
Construction & Preparation queries & Base-estimator queries\\
\midrule
AGGW~\cite{AGGW} with upfront amplification & $O(p^{-1/2})$ & $O(p^{-1/2}\log(2/p))$\\
\addlinespace
Theorem~\ref{thm:main} & $O(p^{-1/2})$ & $O(p^{-1/2})$\\
\midrule
Chen--de Wolf~\cite{CDW} with upfront amplification & --- & $O(\sqrt N\log(2N))$\\
\addlinespace
Gao--Ji--Wang~\cite{GJW}, $k=1$, with upfront amplification & --- & $O(\sqrt N\log(2N))$\\
\addlinespace
Corollary~\ref{cor:uniform} & --- & $O(\sqrt N)$\\
\bottomrule
\end{tabular}
\end{table}

Table~\ref{tab:minimum-comparison} compares our algorithm with the previously discussed constructions of AGGW~\cite{AGGW} and Chen and de Wolf~\cite{CDW}. Gao et al.~\cite{GJW} extend approximate minimum finding to the selection of $k$ distinct labels. In the special case $k=1$, their Algorithm~1 reduces to a single call to AGGW's generalized minimum finder, using uniform preparation over labels followed by the numerical estimator. Appendix~\ref{app:upfront-comparison} derives their query bounds under our oracle assumptions (Definition~\ref{def:oracle}), using confidence amplification before search.

Optimal minimum finding with bounded-error comparisons was obtained by Wang and Ying~\cite{WY}. Allcock et al.~\cite{ABBLS} give query and time bounds for search and minimization with subroutines that return a specified integer answer for each input with bounded error. The non-Boolean purification result of Belovs and Jeffery~\cite[Section~7]{BJ} also assumes that one particular output occurs with probability greater than $1/2$. A numerical estimator may instead distribute its success probability among several values within $\eps$ of the objective, with no individual value occurring with probability greater than $1/2$. Quek et al.~\cite{QCR} consider comparisons that may be unreliable when the values being compared are close. Their result requires a local-density assumption, while our result allows arbitrarily close objective values and arbitrary ties.

Search with neutral inputs~\cite[Section 3.2]{KNW} allows threshold decisions to be ambiguous. Its witness guarantee, however, does not specify how the probability of success is distributed among valid witnesses. For example, an algorithm satisfying this guarantee could always return the largest score below the incumbent. Each update would then reduce the rank only by the weight assigned to that score.

Comparisons with related work on bounded-expectation selection~\cite{WYLC,HHZ} and ground-energy estimation~\cite{JW,MDW,SDW} are given in Section~\ref{sec:applications}.

\subsection{Organization}

Section~\ref{sec:model} formalizes the oracle model and the main results. The proof is developed in Sections~\ref{sec:pairs}--\ref{sec:selection} through validation tests, coherent trials and the adaptive selection procedure. Section~\ref{sec:applications} presents the applications and comparisons with existing methods. Section~\ref{sec:discussion} concludes the paper. The appendices contain supplementary proofs, implementation details and query comparisons for concrete constants.

\section{Model and main results}
\label{sec:model}

We give a more formal statement of the oracle model and main results, and prove lower bounds on the number of estimator queries and the additive selection error. All systems are assumed to be finite-dimensional. We use the Euclidean norm for vectors and the spectral norm for operators. Logarithms are assumed to be natural unless a base is specified. The main parameters used throughout the paper are listed in Table~\ref{tab:notation}.

\begin{table}[htbp]
\centering\small
\caption{Parameters used throughout the paper.}
\label{tab:notation}
\begin{tabular}{@{}lp{.78\textwidth}@{}}
\toprule
Symbol & Meaning\\
\midrule
$p,a_p$ & Known probability level $p$ and corresponding quantile $a_p$, which is used only in the analysis\\
\addlinespace
$\eps$ & Additive accuracy of the numerical estimator\\
\addlinespace
$\delta,u$ & Upper bound $\delta$ on the overall failure probability and $u=\log(2/\delta)$\\
\addlinespace
$s_P,w$ & Numbers of qubits used for preparation and in the estimator workspace, respectively\\
\addlinespace
$v,L,U$ & Returned estimate $v$ and endpoints $L=v-\eps$ and $U=v+\eps$ of the reported interval\\
\addlinespace
$\tau,R_j(\tau),K$ & Current estimate $\tau$, total weight $R_j(\tau)$ of acceptable pairs with estimates below $\tau$ at depth $j$, and maximum recursion depth $K$\\
\addlinespace
$\eta$ & Required additive accuracy in the applications\\
\addlinespace
$\omega$ & Lower bound on the amplitude overlap of the guiding state with the ground space, giving $p=\omega^2$\\
\bottomrule
\end{tabular}
\end{table}

\subsection{Oracle model}

Let $\mathcal C$ be the candidate register and $\mathcal R$ the auxiliary registers of the preparation unitary $P$. We consider a target of the form \[ F=\sum_{i\in I} f_i\Pi_i, \qquad 0\le f_i\le1, \] where $I$ is finite and $\{\Pi_i\}$ is a family of mutually orthogonal projectors summing to the identity on $\mathcal C$. Projectors can have rank greater than one. Classical labels correspond to the special case where $\Pi_i=\ket i\bra i$. We assume access to the objective via the estimator below. We are not given the projectors, a method for preparation of states in different sectors, or an independent oracle for evaluation or measurement of $F$.

Let $P$ be a unitary that prepares the state
\[
 \ket\Omega_{\mathcal C\mathcal R}=P\ket0.
\]
We denote by $\mu_i$ the probability of obtaining sector $i$ in a hypothetical measurement of this state with the projectors $\{\Pi_i\}$, so that
\[
 \mu_i=\norm{(\Pi_i\otimes I_{\mathcal R})\ket\Omega}^2.
\]
For a set $E$ of objective values, we write $\mu\{f\in E\}=\sum_{i:f_i\in E}\mu_i$ for the probability that the measured value belongs to $E$.

\begin{definition}
\label{def:oracle}
A \emph{coherent numerical estimator}, also called a \emph{base estimator}, is a unitary on the candidate register $\mathcal C$ and a work register $\mathcal W$ of the form
\begin{equation}
 A=\sum_{i\in I}\Pi_i\otimes A_i.
 \label{eq:nd}
\end{equation}
It acts as the identity on $\mathcal R$ and on all previously retained work registers.

The estimator has a designated output register and a known decoding function. For each $i$, measuring this register in the computational basis in the state $A_i\ket0_{\mathcal W}$ and decoding the outcome gives a finite-valued random variable $Y_i\in[0,1]$ satisfying
\begin{equation}
 \Pr\{|Y_i-f_i|\le\eps\}\ge2/3
 \quad\text{for every }i.
 \label{eq:accuracy}
\end{equation}
\end{definition}
The decoded estimates can be clipped to $[0,1]$ without increasing their error, because the objective values lie in this interval.

If the candidate is in sector $i$, $A$ acts only on the estimator workspace. For any state $\ket{\psi_i}_{\mathcal C\mathcal R}$ supported in sector $i$,
\[
 A\bigl(\ket{\psi_i}_{\mathcal C\mathcal R}\ket0_{\mathcal W}\bigr)
 =\ket{\psi_i}_{\mathcal C\mathcal R}\otimes A_i\ket0_{\mathcal W}.
\]
Since $A_i\ket{0}_{\mathcal W}$ does not depend on the candidate state within sector $i$, the estimate has the distribution of $Y_i$ for every such state. For a fixed sector $i$, independent estimates are obtained by resetting the workspace to $\ket0_{\mathcal W}$ between successive applications of $A$. In case of phase estimation, the evolution multiplies all states in a given energy eigenspace by the same phase. The energy eigenspaces can therefore be used as the sectors in~\eqref{eq:nd}.

We assume that $P$, $A$ and their inverses are available. The preparation query count $Q_P$ is the total number of applications of $P$ and $P^\dagger$, and the base-estimator query count $Q_A$ is the total number of applications of $A$ and $A^\dagger$. Let $s_P$ denote the number of qubits in the preparation register $\mathcal C\mathcal R$, and let $w$ denote the number of qubits in the estimator workspace $\mathcal W$, including the numerical output register and all registers used for coherent decoding. Known rotations are treated as exact in the main proof. Gate approximations are treated in Appendix~\ref{app:precision}, where it is shown that the change in success probability is bounded by the sum of their operator-norm errors.

In case the output is a retained candidate, let $X_{\rm out}$ denote the objective value that would have been obtained by a final measurement of $F$. Probabilities involving $X_{\rm out}$ refer to the hypothetical experiment of running the algorithm and then making this measurement.

\subsection{Main theorem}

For analysis define
\begin{equation}
 a_p=\min\{a:\mu\{f\le a\}\ge p\}.
 \label{eq:quantile}
\end{equation}
Finiteness implies $\mu\{f<a_p\}<p$. The algorithm is not given $a_p$.

\begin{theorem}
\label{thm:main}
Under Definition~\ref{def:oracle}, for every known $0<p\le1$, $\eps>0$, and $0<\delta<1/2$, an algorithm returns a candidate and real numbers $v,L,U$, with $L=v-\eps$ and $U=v+\eps$ on every execution, such that jointly with probability at least $1-\delta$,
\begin{equation}
 |v-X_{\rm out}|\le\eps,\qquad v\le a_p+\eps.
 \label{eq:main-success}
\end{equation}
Writing $u=\log(2/\delta)$, its worst-case resource bounds are
\begin{equation}
 Q_P=O(p^{-1/2}u),\quad
 Q_A=O(p^{-1/2}u+u^2),\quad
 S_Q=O(s_P+w+\log\log(4/p)).
 \label{eq:main-cost}
\end{equation}
The constants are universal and independent of the number of possible estimator outputs. The algorithm knows neither $\mu$ nor a qualifying objective threshold. It uses no controlled $P$ or controlled $A$.
\end{theorem}
\label{rem:output}
Theorem~\ref{thm:main} gives
\begin{equation}
 X_{\rm out}\in[L,U],\qquad
 \mu\{f<L\}<p,\qquad U-L=2\eps
 \label{eq:certificate}
\end{equation}
with probability at least $1-\delta$, where $L\le a_p$ gives a bound on the probability mass below $L$. For every $a$ with $\mu\{f\le a\}\ge p$, the selected objective value therefore satisfies
\begin{equation}
 X_{\rm out}\le a+2\eps.
 \label{eq:selection}
\end{equation}
Let $\bar\rho$ denote the density matrix of the returned candidate register $\mathcal C$, averaged over the algorithm's measurement outcomes and classical random choices. For every $a$ with $\mu\{f\le a\}\ge p$, measuring $F$ on this register gives
\[
 \Pr\{X_{\rm out}\le a+2\eps\}
 =\Tr\!\left(\bar\rho\sum_{i:f_i\le a+2\eps}\Pi_i\right)
 \ge1-\delta.
\]

The selection and validation procedures do not use $\eps$, assuming that the base estimator is given. Its value is needed only to form $L$ and $U$.

For uniform preparation over $N$ classical labels, choosing $p=1/N$ gives the following corollary.

\begin{corollary}
\label{cor:uniform}
For $N$ classical labels with known uniform preparation, let $A$ be a label-preserving estimator satisfying Definition~\ref{def:oracle}. With probability at least $1-\delta$, an algorithm returns a label $s$ and numbers $v,L,U$ satisfying the output guarantee of Theorem~\ref{thm:main} with $p=1/N$; in particular,
\[
 f(s)\le\min_i f(i)+2\eps.
\]
It uses $O(\sqrt N\,u+u^2)$ base-estimator queries and $O(w+\log(2N))$ live quantum qubits, where $u=\log(2/\delta)$.
\end{corollary}
\begin{proof}
With uniform preparation, each label has probability $1/N$, so every minimum-value block has prior mass at least $1/N$. The query bound follows by applying Theorem~\ref{thm:main} with $p=1/N$. Uniform preparation requires $O(\log(2N))$ qubits. Since $p=1/N$, the additional term $\log\log(4/p)=\log\log(4N)$ is also bounded by $O(\log(2N))$. The classical output is obtained by measuring the candidate register in the label basis.
\end{proof}

\subsection{Lower bounds}

The next two propositions show that both the additive-error constant $2$ and the base-estimator query scaling in Theorem \ref{thm:main} are individually optimal.
\begin{proposition}
\label{prop:accuracy-lower}
Let $0<\eps\le1/2$. Under the numerical-estimation promise alone, no algorithm can guarantee additive selection error below $2\eps$ with success probability greater than $1/2$ on every instance, regardless of the number of queries.
\end{proposition}

\begin{proof}
Take two labels with uniform preparation, and fix an estimator unitary that returns $1/2$ on both labels. This estimator satisfies the numerical promise for either of the objectives
\[
(f(0),f(1))=
\left(\frac12-\eps,\frac12+\eps\right)
\quad\text{and}\quad
(f(0),f(1))=
\left(\frac12+\eps,\frac12-\eps\right).
\]
Since the algorithm has no other access to $f$, its output distribution is the same in both cases.

For $p=1/2$ and $a=1/2-\eps$, selection error below $2\eps$ requires label $0$ in the first case and label $1$ in the second. Let $r$ be the probability of obtaining label $0$ from the output, measuring in the label basis if necessary. The two success probabilities are therefore $r$ and $1-r$, so at least one is at most $1/2$.
\end{proof}

\begin{proposition}
\label{prop:search-lower}
Suppose $2\eps<1$. At constant confidence, the $\Theta(p^{-1/2})$ dependence of the base-estimator query count is optimal in the worst case. This holds even for $p=1/N$ with classical labels and free, known uniform preparation.
\end{proposition}

\begin{proof}
Consider unstructured search on $N$ labels with a unique marked label $z$. Set $f(z)=0$ and $f(i)=1$ for $i\ne z$, and use the exact value oracle for $f$ as the numerical estimator. Each estimator query can be simulated with a constant number of search-oracle queries.

With $p=1/N$ and $a=0$, a successful output must satisfy $f(s)\le2\eps<1$, so $s=z$. The algorithm therefore finds the marked label with the same success probability, measuring its output in the label basis if necessary.

The quantum search lower bound~\cite{BBBV,BBHT} then implies $\Omega(\sqrt N)=\Omega(p^{-1/2})$ base-estimator queries. Corollary~\ref{cor:uniform} gives the matching upper bound at constant confidence.
\end{proof}

\section{Validation of numerical estimates}\label{sec:pairs}

To validate an estimate $y$, the algorithm uses new estimates obtained from the same candidate. For a fixed sector $i$, the estimate $y$ must be $\eps$-accurate if both $\Pr\{Y_i\le y\}$ and $\Pr\{Y_i\ge y\}$ exceed $1/3$. This section develops validation tests based on these two probabilities.

\subsection{Accuracy criteria}

Let $A_{\rm sample}$ denote an application of $A$ to a fresh estimator workspace. Preparing a candidate and estimating its value gives
\[
 B_{\rm prep}=A_{\rm sample}(P\otimes I),\qquad
 \pi_{i,y}=\mu_i\Pr\{Y_i=y\},
\]
where $\pi_{i,y}$ is the joint probability of sector $i$ and decoded estimate $y$. We call the stored numerical estimate $y$ the score.

For a pair $z=(i,y)$, the probabilities that a new estimate is at most or at least $y$ are
\begin{equation}
 q_i^-(y)=\Pr\{Y_i\le y\},\qquad
 q_i^+(y)=\Pr\{Y_i\ge y\}.
 \label{eq:cdfs}
\end{equation}
Choose thresholds satisfying
\begin{equation}
 \frac13<\gamma<\alpha<\beta<\frac12,
 \label{eq:margins}
\end{equation}
and keep them fixed throughout each run. Define the good, bad, and acceptable pairs, respectively, by
\begin{equation}
 G=\{z:q_i^-(y),q_i^+(y)\ge\beta\},\quad
 B=\{z:\min(q_i^-(y),q_i^+(y))\le\alpha\},\quad
 S=B^c.
 \label{eq:classes}
\end{equation}
It follows that every good pair is acceptable. Furthermore, every acceptable pair must satisfy $|y-f_i|\le\eps$. To see this, observe that if $y<f_i-\eps$, then $Y_i\le y$ can occur only when the new estimate is inaccurate, so $q_i^-(y)\le1/3$. Similarly, $y>f_i+\eps$ implies that $q_i^+(y)\le1/3$. Neither of these cases can occur for a pair in $S$, because $\alpha>1/3$.

The restriction $\beta<1/2$ ensures that good pairs have probability at least $1-2\beta>0$ in every sector.

The thresholds $\alpha$ and $\beta$ are fixed constants independent of $p$ and $\delta$.

\begin{lemma}\label{lem:mass}
The probability of preparing a good pair with score at most $a_p+\eps$ is at least $(1-2\beta)p$:
\[
 \pi\{z\in G:y\le a_p+\eps\}\ge(1-2\beta)p.
\]
\end{lemma}

\begin{proof}
Fix a sector $i$, and let $F_i$ be the CDF of $Y_i$. After sampling $Y_i=y$, sample $U$ uniformly from $[F_i(y^-),F_i(y)]$. The resulting random variable $U$ is uniform on $[0,1]$, and
\[
 q_i^-(Y_i)=F_i(Y_i)\ge U,
 \qquad
 q_i^+(Y_i)=1-F_i(Y_i^-)\ge 1-U.
\]
Since $U$ is uniform on $[0,1]$, it lies in $[\beta,1-\beta]$ with probability $1-2\beta$. In that interval, $q_i^-(Y_i)\ge U\ge\beta$ and $q_i^+(Y_i)\ge1-U\ge\beta$, so $(i,Y_i)$ is a good pair.

Therefore, since every good pair is accurate, if $f_i\le a_p$, the same event also gives
\[
 Y_i\le f_i+\eps\le a_p+\eps.
\]
The preparation assigns total mass at least $p$ to sectors with $f_i\le a_p$. Hence
\[
 \pi\{z\in G:y\le a_p+\eps\}
 \ge (1-2\beta)\sum_{i:f_i\le a_p}\mu_i
 \ge (1-2\beta)p.
\]
\end{proof}

\subsection{Validation by repeated estimation}

\begin{lemma}\label{lem:validation}
For fixed $0<l<h<1$ and $0<\zeta<1/2$, there is a validation procedure $\Val_{l,h}(y,\zeta)$ that uses
\begin{equation}
 n_{l,h}(\zeta)
 =\left\lceil\frac{2}{(h-l)^2}\log\frac2\zeta\right\rceil
 \label{eq:validation-count}
\end{equation}
forward base-estimator queries and one reusable estimator workspace. For every sector $i$ and stored score $y$, it satisfies
\begin{align}
 q_i^-(y),q_i^+(y)\ge h
 &\Longrightarrow \Pr\{\mathrm{accept}\mid i,y\}\ge1-\zeta,
 \label{eq:val-complete}\\
 \min(q_i^-(y),q_i^+(y))\le l
 &\Longrightarrow \Pr\{\mathrm{accept}\mid i,y\}\le\zeta.
 \label{eq:val-sound}
\end{align}
These bounds also hold conditional on earlier classical history of non-zero probability.
\end{lemma}

\begin{proof}
Run the estimator $n=n_{l,h}(\zeta)$ times to obtain independent estimates $Y_i^{(1)},\ldots,Y_i^{(n)}$. Define
\[
 \widehat q^-=\frac1n\sum_{j=1}^n
 \mathbf{1}_{\{Y_i^{(j)}\le y\}},
 \qquad
 \widehat q^+=\frac1n\sum_{j=1}^n
 \mathbf{1}_{\{Y_i^{(j)}\ge y\}}.
\]
The procedure $\Val_{l,h}(y,\zeta)$ accepts the stored score $y$ if both $\widehat q^-$ and $\widehat q^+$ are at least $(l+h)/2$, and rejects it otherwise.

By Definition~\ref{def:oracle}, repeated calls to the estimator produce independent outputs with the same distribution as $Y_i$, conditional on the sector and earlier classical history. If both $q_i^-(y)$ and $q_i^+(y)$ are at least $h$, Hoeffding's inequality gives
\[
 \Pr\!\left\{\widehat q^\pm<\frac{l+h}{2}\,\middle|\,i,y\right\}
 \le \exp\!\left(-\frac{n(h-l)^2}{2}\right)
 \le \frac{\zeta}{2}.
\]
Rejection requires at least one of these two events to occur, therefore its probability is at most $\zeta$.

For soundness, suppose first that $q_i^-(y)\le l$. Acceptance requires $\widehat q^-\ge(l+h)/2$, and hence
\[
 \widehat q^- - q_i^-(y)
 \ge \frac{l+h}{2}-q_i^-(y)
 \ge \frac{h-l}{2}.
\]
The indicators defining $\widehat q^-$ are independent Bernoulli random variables with mean $q_i^-(y)$. Hoeffding's inequality therefore gives
\[
 \Pr\{\mathrm{accept}\mid i,y\}
 \le
 \Pr\!\left\{
   \widehat q^- - q_i^-(y)\ge\frac{h-l}{2}
   \,\middle|\,i,y
 \right\}
 \le
 \exp\!\left(-\frac{n(h-l)^2}{2}\right)
 \le \frac{\zeta}{2}
 \le \zeta.
\]
Here the earlier classical history is fixed and suppressed in the conditioning notation. If $q_i^+(y)\le l$, the same argument applies to $\widehat q^+$. Since $\min(q_i^-(y),q_i^+(y))\le l$ implies that at least one of these two cases holds, this proves \eqref{eq:val-sound}, also conditional on any earlier classical history of positive probability.

The union bound on rejection holds when $\widehat q^-$ and $\widehat q^+$ are computed from the same estimates. The numbers of outputs at most $y$ and at least $y$ are stored classically. After each query, the estimator workspace is measured and reset so that it can be reused.

\end{proof}

During selection, \emph{inner validation} checks proposed replacements, using thresholds $(l,h)=(\gamma,\alpha)$. During confidence amplification, \emph{outer validation} checks candidates returned by selection runs, using thresholds $(l,h)=(1/3,\gamma)$.

Inner validation accepts a pair with $\min(q_i^-(y),q_i^+(y))\le\gamma$ with probability at most $\zeta$. For every remaining pair, both probabilities exceed $\gamma>1/3$, which implies $|y-f_i|\le\eps$. These pairs also satisfy the condition under which outer validation accepts with probability at least $1-\zeta$.

The preceding lemma bounded the probabilities of false acceptance and false rejection by the same parameter $\zeta$. The following lemma allows these probabilities to be bounded separately, which can reduce the number of estimator queries needed for validation.

\begin{lemma}\label{lem:asymmetric-validation}
Let $0<l<h<1$, $0<\zeta<1$, and $0<v_{\rm acc}<1$. Run the estimator $n$ times. Accept the stored score $y$ if at least $k$ outputs are at most $y$ and at least $k$ outputs are at least $y$.

If
\begin{equation}
 \Pr\{\operatorname{Bin}(n,l)\ge k\}\le\zeta,
 \qquad
 \Pr\{\operatorname{Bin}(n,h)<k\}\le\frac{1-v_{\rm acc}}{2},
 \label{eq:asymmetric-validation}
\end{equation}
then the procedure accepts with probability at most $\zeta$ when $\min(q_i^-(y),q_i^+(y))\le l$, and with probability at least $v_{\rm acc}$ when both probabilities are at least $h$.

These bounds hold in every sector after conditioning on the measurement outcomes and classical random choices before validation.
\end{lemma}

\begin{proof}
Fix the sector $i$ and the stored score $y$. By Definition~\ref{def:oracle}, the estimates collected during validation are independent with distribution $Y_i$, regardless of the measurement outcomes and classical random choices before validation.

Let $N^-$ be the number of the estimator outputs at most $y$, and let $N^+$ be the number of estimator outputs at least $y$. By Definition~\ref{def:oracle},
\[
 N^-\sim\operatorname{Bin}(n,q_i^-(y)),
 \qquad
 N^+\sim\operatorname{Bin}(n,q_i^+(y)).
\]

If $q_i^-(y)\le l$, then
\[
 \Pr\{\mathrm{accept}\}
 \le \Pr\{N^-\ge k\}
 \le \Pr\{\operatorname{Bin}(n,l)\ge k\}
 \le\zeta.
\]
The second inequality holds because the probability of at least $k$ successes increases with the success probability. The same argument applies if $q_i^+(y)\le l$.

If both probabilities are at least $h$, then
\[
 \Pr\{\mathrm{reject}\}
 \le \Pr\{N^-<k\}+\Pr\{N^+<k\}
 \le 2\Pr\{\operatorname{Bin}(n,h)<k\}
 \le 1-v_{\rm acc}.
\]
Both counts use the same estimator outputs. They are not required to be independent, since the bound on the rejection probability follows from a union bound.
\end{proof}

The same guarantees follow from Lemma~\ref{lem:validation} by choosing its error parameter $\eta$ so that
\[
0<\eta<\frac12,
\qquad
\eta\le\min\{\zeta,1-v_{\rm acc}\}.
\]
Choosing $n$ and $k$ directly from the two binomial inequalities in~\eqref{eq:asymmetric-validation} can reduce the number of estimator queries, as illustrated in Appendix~\ref{app:concrete-schedule}.

\section{Coherent proposal generation}
\label{sec:trials}

Given a threshold $\tau$, we construct a trial that seeks a candidate with a stored estimate $y<\tau$. The trial uses coherent filters based on the probabilities in~\eqref{eq:cdfs} to suppress bad pairs and interleaves these filters with amplitude amplification to increase the probability of obtaining an acceptable pair. We also derive bounds on the number of queries and qubits used by each trial.

\subsection{Polynomial filters}
\label{sec:filters}

For each filter, we give a lower bound on the probability of accepting a good pair and an upper bound on the probability of accepting a bad pair. These bounds depend on the thresholds $\alpha,\beta$ in~\eqref{eq:margins} and on an error parameter for each filter.

The initialized block of a unitary is obtained by projecting its work registers onto the all-zero state at both the input and output. We also call this the clean block.

The estimator workspace containing the stored estimate is written $YG_0$, where $Y$ contains the encoded estimate and $G_0$ the remaining work registers. These registers are kept coherent during the trials. Let $Q_y$ project onto the basis states of $Y$ that decode to $y$, and define
\[
\widetilde\Pi_{i,y}=\Pi_i\otimes Q_y.
\]
For either comparison $Y_i\le y$ or $Y_i\ge y$, let $E_y$ project onto the estimator-output subspace whose decoded values satisfy that comparison. The Hermitian unitary
\[
 J_y=A^\dagger(I-2E_y)A
\]
has initialized block $1-2q_i^\pm(y)$. Put
\begin{equation}
 a_{\rm sel}=\frac1{2-\alpha-\beta},\qquad
 x(q)=\frac{\alpha+\beta-2q}{2-\alpha-\beta},\qquad
 g=\frac{\beta-\alpha}{2-\alpha-\beta}>0.
 \label{eq:signal-block}
\end{equation}
On a selector qubit, prepare $V_0\ket0=\sqrt{a_{\rm sel}}\ket0+\sqrt{1-a_{\rm sel}}\ket1$ and define
\begin{equation}
 U_y=(V_0^\dagger\otimes I)
 (\ket0\bra0\otimes J_y-\ket1\bra1\otimes I)
 (V_0\otimes I).
 \label{eq:shifted-signal}
\end{equation}
Its initialized block is $X=\sum_{i,y}x(q_i^\pm(y))\widetilde\Pi_{i,y}$. The range of $x$ lies in $[-1,1]$, and
\begin{equation}
 q\le\alpha\Longrightarrow x(q)\ge g,\qquad
 q\ge\beta\Longrightarrow x(q)\le-g.
 \label{eq:signal-gap}
\end{equation}

For each level $\ell$, choose $0<u_\ell<1$ and an odd real polynomial $\varphi_\ell$ of degree $d_\ell$, bounded by one on $[-1,1]$, such that
\begin{equation}
 |\varphi_\ell(x)-\operatorname{sign}(x)|\le u_\ell
 \quad\text{for }g\le|x|\le1.
 \label{eq:filter-polynomial-contract}
\end{equation}
Appendix~\ref{app:polynomial} constructs such polynomials with $d_\ell=O(g^{-1}\log(2/u_\ell))$. Thus a geometrically decreasing $u_\ell$ gives $d_\ell=O(\ell+1)$ for fixed margins. We assume this degree bound when deriving the uniform asymptotic resource bounds.

\begin{lemma}\label{lem:clean}
For every $\ell\ge0$, there is a full unitary $T_\ell$ on $\mathcal C Y$ and a common $O(w)$-qubit work register $W_0$ whose initialized block is
\begin{equation}
 (I\otimes\bra0_{W_0})T_\ell(I\otimes\ket0_{W_0})
 =\sum_z k_\ell(z)\widetilde\Pi_z,\qquad 0\le k_\ell(z)\le1.
 \label{eq:clean-block}
\end{equation}
Its scalar acceptance probability $e_\ell(z)=k_\ell(z)^2$ obeys
\begin{equation}
 e_\ell(z)\ge(1-u_\ell/2)^4\quad(z\in G),\qquad
 e_\ell(z)\le(u_\ell/2)^2\quad(z\in B).
 \label{eq:filter-bounds}
\end{equation}
The filter, its inverse, and controlled versions use at most $4d_\ell$ base-estimator queries, no preparation queries, and no controlled unknown oracle. They are independent of the incumbent threshold and act as the identity on preparation references and stored sampling garbage.
\end{lemma}
\begin{proof}
Real-polynomial QSVT~\cite[Corollary 18]{GSLW} applied to $U_y$ produces a unitary $S_\ell$ with clean block $\varphi_\ell(X)$, using $d_\ell$ signal traversals. For a Hermitian block, an odd singular-value transformation equals the corresponding signed eigenvalue polynomial. The real-polynomial selector combines opposite phase sequences using the same number of traversals and a constant number of extra qubits.

On an additional selector qubit $h$, the unitary $X_hH_h\operatorname{ctrl}_h(S_\ell)H_h$ has clean block
\[
 k_\ell^\pm(i,y)=\frac{1-\varphi_\ell(x(q_i^\pm(y)))}2\in[0,1].
\]
Use separate work for the two comparisons and multiply these blocks. Each is scalar on a sector/score block and acts trivially on the other's work, so $k_\ell=k_\ell^-k_\ell^+$. On $G$, both factors are at least $1-u_\ell/2$. On $B$, at least one is at most $u_\ell/2$. Squaring gives~\eqref{eq:filter-bounds}.

One signal traversal uses two base-estimator queries. Controlled traversals require no controlled estimator, since for every known control condition $b$,
\begin{equation}
 \operatorname{ctrl}_b J_y
 =A^\dagger\operatorname{ctrl}_b(I-2E_y)A.
 \label{eq:controlled-signal}
\end{equation}
The oracle operations cancel on inactive branches over the full workspace. Both base-estimator queries are still counted. The selector preparation and unpreparation also cancel there. Thus each comparison uses $2d_\ell$ base-estimator queries, giving $4d_\ell$ in total, including for controlled and inverse filters. Acceptance tests all work, including the selectors. The phase list is classical; quantum width is independent of $\ell$.
\end{proof}

For subsequent bounds, choose any certified numbers
\begin{equation}
 0<\mathsf r_\ell\le(1-u_\ell/2)^4,\qquad
 (u_\ell/2)^2\le\mathsf b_\ell\le1.
 \label{eq:filter-certificates}
\end{equation}
Then $e_\ell\ge\mathsf r_\ell$ on $G$ and $e_\ell\le\mathsf b_\ell$ on $B$. These certificates can use either analytic error bounds or exact polynomial integration.

The threshold $\tau$ can be imposed by appending a flag qubit $h_\tau$ to $W_0$. Define $\widehat T_\ell(\tau)$ by applying $T_\ell$ and flipping $h_\tau$ when $y\ge\tau$. Its clean block on $z$ is
\[
 k_\ell(z)\one_{\{y<\tau\}}.
\]
Write $W=W_0h_\tau$ for the common workspace used at every recursive level.

\begin{remark}
\label{rem:exact-gates}
We use exact known gates, as permitted by the query model in Section~\ref{sec:model}. Appendix~\ref{app:precision} bounds the effect of approximating these gates on the full computation.
\end{remark}

\subsection{Amplification with reusable workspace}\label{sec:recursion}

We combine the filters with amplitude amplification to construct recursive trials that accept only estimates below $\tau$. Fix a maximum depth $K\ge1$, and introduce a counter register $D$ with labels $0,\ldots,K$. The label $0$ denotes the active component and the other labels identify the first filter that rejected a component. Subsequent filters act only on the active component. The filter workspace is in the zero state on this component and can therefore be reused at every level. Let
\[
\Pi=\ket0\bra0_D\otimes I
\]
be the projector onto the active subspace.

For $0\le\ell<K$, define the unitary $F_\ell(\tau)$ by applying $\widehat T_\ell(\tau)$ controlled on $D=0$, followed by swapping counter values $0$ and $\ell+1$ conditional on $W\ne0$.

Let $R_0=I-2\ket0\bra0$ be the reflection about the all-zero state of all trial registers. This reflection and comparisons on $D$ use only known gates. Define
\begin{align}
 B_1(\tau)&=F_0(\tau)B_{\rm prep},\label{eq:recursion-base}\\
 B_{j+1}(\tau)&=F_j(\tau)
 \bigl[-B_j(\tau)R_0B_j(\tau)^\dagger(I-2\Pi)\bigr]B_j(\tau),
 \qquad 1\le j<K.
 \label{eq:recursion}
\end{align}
The bracketed operator performs one amplitude-amplification step on the active component before the next filter is applied.

When the new filter at level $\ell$ is applied, only counter labels $0,\ldots,\ell$ are occupied, and the active component has $W=0$. For components rejected at this level, the swap changes the counter label from $0$ to $\ell+1$. The active component can reuse the filter workspace at the next level.

A trial of depth $j$ applies $B_j(\tau)$ to the all-zero state and tests whether $D=0$. If so, it measures the stored score.

\subsection{Proposal probabilities}
\label{sec:proposal-law}

\begin{lemma}\label{lem:recursion}
For every threshold $\tau$, write $B_j(\tau)\ket0=g_j+\chi_j$, where $g_j=\Pi B_j(\tau)\ket0$ and $\Pi\chi_j=0$. The active workspace is zero, and $\chi_j$ uses only counter values $1,\ldots,j$. Define
\begin{equation}
 \omega_j(z)=\pi_z\prod_{\ell=0}^{j-1}e_\ell(z),\quad
 R_j(\tau)=\sum_{z\in S:y<\tau}\omega_j(z),\quad
 H_j(\tau)=\sum_{z\in B:y<\tau}\omega_j(z).
 \label{eq:rank-weights}
\end{equation}
There are real numbers $c_j(\tau)$ such that
\begin{equation}
 \Pr\{\text{active and pair }z\}
 =c_j(\tau)^2\omega_j(z)\one_{\{y<\tau\}},\qquad
 s_j:=\norm{g_j}^2=c_j^2(R_j+H_j),
 \label{eq:proposal-law}
\end{equation}
with $c_1=1$, $c_{j+1}=(3-4s_j)c_j$, and $|c_j|\le3^{j-1}$. The preparation cost is exactly $3^{j-1}$, and the estimator cost $q^A_j$ obeys
\begin{equation}
 q^A_1=1+4d_0,\qquad q^A_{j+1}=3q^A_j+4d_j,
 \qquad q^A_j=O(3^{j-1}).
 \label{eq:trial-cost}
\end{equation}
The circuit $B_j(\tau)$ uses $O(s_P+w+\log(K+1))$ qubits.
\end{lemma}
\begin{proof}
The first filter leaves active amplitude $k_0(z)$ times the prepared pair amplitude, with all its work zero. It records rejection in label $1$. For an arbitrary normalized $g+\chi$ with $g$ active and $s=\norm g^2$, the expression in square brackets in~\eqref{eq:recursion} maps
\[
 g+\chi\ \longmapsto\ (3-4s)g+(1-4s)\chi.
\]
This follows by expanding the reflection $B_jR_0B_j^\dagger$ about $B_j\ket0$. Applying $F_j$ then multiplies each active pair amplitude by $k_j(z)$ and sends newly rejected work to the unused counter label $j+1$. Older rejected counter values are unchanged. Induction proves the active workspace invariant and~\eqref{eq:proposal-law}. Repeated threshold indicators reduce to the same strict indicator, and $|3-4s_j|\le3$.

The recursion for \(B_{j+1}\) uses two calls to \(B_j\), one to its inverse \(B_j^\dagger\), and an additional filter. Thus the query recurrences are exact for the indicated implementation, or upper bounds when a filter uses fewer traversals. Since $d_\ell=O(\ell+1)$, the geometric sum gives
\[
 \frac{q^A_j}{3^{j-1}}
 =1+\sum_{\ell<j}\frac{4d_\ell}{3^\ell}=O(1).
\]
All three recursive calls act sequentially on the same registers. The recursion can be unrolled classically into a gate sequence, without requiring a quantum register to keep track of the recursive calls. Rejection labels require only $\lceil\log_2(K+1)\rceil$ qubits, and all filter work has width $O(w)$.
\end{proof}

The nonnegative weights $\omega_j$ depend on the level but not on $\tau$. They are nonincreasing with $j$ on every pair, and $R_j$ is nondecreasing in $\tau$. Thus an accepted update makes every $R_j(\tau)$ nonincreasing. The score can decrease without a strict decrease in acceptable rank, because an accepted pair can lie outside $S$. Define the certified retention and amplified leakage at depth $j$ by
\begin{equation}
 r_*(j)=\prod_{\ell=0}^{j-1}\mathsf r_\ell,\qquad
 \Lambda_j=9^{j-1}\prod_{\ell=0}^{j-1}\mathsf b_\ell.
 \label{eq:retention-leakage}
\end{equation}
Then
\begin{equation}
 \omega_j(z)\ge r_*(j)\pi_z\quad(z\in G),\qquad
 9^{j-1}H_j(\tau)\le\Lambda_j.
 \label{eq:retention}
\end{equation}
Thus $r_*(j)$ lower-bounds the fraction of each good pair's initial weight remaining after filtering, and $\Lambda_j$ upper-bounds the active probability contributed by unacceptable pairs after amplification. For a uniform asymptotic schedule we will certify constants $r_*>0$ and $\Lambda\ge0$ satisfying
\begin{equation}
 r_*(j)\ge r_*\quad(j\ge1),\qquad
 \sum_{j\ge1}\Lambda_j\le\Lambda.
 \label{eq:leakage}
\end{equation}
These quantities retain the dependence on the filter choices. Appendix~\ref{app:filter-feasibility} proves feasibility of the uniform conditions. Appendix~\ref{app:concrete-schedule} gives finite numerical schedules using the same filter certificates.

\section{Adaptive selection}
\label{sec:selection}

We first give conditions on the schedule, then describe the selection algorithm and state bounds on its success probability and query cost.  A geometric schedule proves the asymptotic result. The schedule and inner validation procedure used in the numerical comparisons are specified in Appendix~\ref{app:concrete-schedule}.

\subsection{The selection algorithm}\label{sec:controller}

The selection algorithm keeps the lowest score that passes validation, together with its candidate. Its schedule consists of a depth $K$, trial counts $t_1,\ldots,t_K$, and an inner validation procedure $\mathcal V$. We give sufficient conditions for these choices before describing the algorithm.

Fix a depth $K\ge1$, cap scales $h_j^\star>0$, and validation completeness $0<v_{\rm acc}<1$. The cap scales set the target rank bounds $R_j(\tau)\le h_j^\star9^{-(j-1)}$. Choose failure allowances $\varepsilon_{\rm cap}, \varepsilon_{\rm val}>0$ with
\[
 \sigma=\varepsilon_{\rm cap}+\varepsilon_{\rm val}<1/2.
\]
With $\Lambda_j$ from~\eqref{eq:retention-leakage}, define
\begin{equation}
 a_j^\star=\prod_{\ell=1}^{j-1}
 \left[1-\frac43(h_\ell^\star+\Lambda_\ell)\right],\qquad
 \eta_j^\star=v_{\rm acc}(a_j^\star)^2h_j^\star.
 \label{eq:finite-gain}
\end{equation}
The empty product is one. Require every factor to be positive and choose certified lower bounds $0<\underline\eta_j\le\eta_j^\star$ with $\underline\eta_j<1$. Choose positive integer trial counts $t_j$ such that
\begin{equation}
 h_K^\star9^{-(K-1)}<r_*(K)(1-2\beta)p,\qquad
 \sum_{j=1}^K(1-\underline\eta_j)^{t_j}\le\varepsilon_{\rm cap},
 \qquad T=\sum_{j=1}^K t_j.
 \label{eq:finite-certificate}
\end{equation}
Use an inner validation procedure $\mathcal V$ with one reusable estimator workspace, classical counts, and at most $n$ estimates per invocation. Require conditional completeness at least $v_{\rm acc}$ on $S$. When $\min(q_i^-,q_i^+)\le\gamma$, require conditional soundness at most
\begin{equation}
 \zeta=\varepsilon_{\rm val}/T
 \label{eq:inner-soundness}
\end{equation}
Lemmas~\ref{lem:validation} and~\ref{lem:asymmetric-validation} provide such validation procedures. One sufficient choice is
\[
 \mathcal V(y)=\Val_{\gamma,\alpha}(y,\min\{\zeta,1-v_{\rm acc}\}).
\]

We now describe the selection algorithm for a schedule satisfying the conditions above. We call this constant-confidence procedure the seed algorithm.
\begin{algorithm}\label{alg:seed}
\leavevmode
\begin{enumerate}[label=\arabic*.,leftmargin=2em,itemsep=3pt]
\item Given $P,A$, their inverses, the decoder, $p$, and a schedule $K,(t_j),\mathcal V$ satisfying the conditions above, initialize $\tau\leftarrow+\infty$ and $\mathsf{inc}\leftarrow\bot$.

\item For each $j=1,\ldots,K$, perform $t_j$ trials as follows.
\begin{enumerate}[label=(\alph*),leftmargin=1.8em,itemsep=2pt]
\item Initialize the trial registers to zero and apply $B_j(\tau)$.
\item Measure $D$. If $D\ne0$, discard the trial and proceed to the next repetition.
\item Measure the stored score $y$. If $y\ge\tau$, discard the trial and proceed to the next repetition. Otherwise, run $\mathcal V$ on the candidate with stored score $y$.
\item If validation accepts, replace the incumbent with this candidate and set $\tau\leftarrow y$. Otherwise, discard the candidate.
\end{enumerate}

\item Return $(\mathsf{inc},\tau)$ if an incumbent was retained, and return $\bot$ otherwise.
\end{enumerate}
\end{algorithm}

\begin{lemma}
\label{lem:latent-sector}
In Algorithm~\ref{alg:seed}, each candidate produced by a completed trial may be assigned a fixed, unobserved sector $i$ without changing the joint distribution of the classical transcript and the final hypothetical objective measurement $X_{\rm out}$. Here the classical transcript consists of the measurement outcomes and classical random choices.

Conditional on $i$ and the transcript before a validation, the estimates used in that validation are independent and have distribution $Y_i$. Thus Lemma~\ref{lem:validation} applies.

The same conclusions hold for the repeated runs and additional validation in Algorithm~\ref{alg:confidence}.
\end{lemma}

\begin{proof}
Fix the classical transcript after a trial has ended and its score has been measured. Let $\rho$ be the conditional state of the candidate register, and set
\[
 p_i=\Tr(\Pi_i\rho).
\]
Consider $n$ subsequent estimator calls, each with its workspace initialized to $\ket0$ and measured afterwards. By~\eqref{eq:nd}, the joint probability of workspace outcomes $x_1,\ldots,x_n$ and sector $i$ in a hypothetical sector measurement after these calls is
\[
 \Pr\{x_1,\ldots,x_n,i\}
 =p_i\prod_{r=1}^n
   \left|\bra{x_r}A_i\ket0\right|^2.
\]
These are the probabilities obtained by choosing $i$ once with probability $p_i$ and then generating independent workspace outcomes according to $A_i\ket0$. Decoding the outcomes gives independent estimates with distribution $Y_i$.

The conditional probabilities may change depending on whether the candidate is accepted or rejected. The next estimator output still has distribution $Y_i$ within a fixed sector. Each new trial uses separate registers and is only dependent on the incumbent through its recorded score. Hence, even if a successful candidate replaces the incumbent, we can use the same calculation after each trial and each validation.

The argument also applies to further validation and selection of candidates returned by independent runs. Since the objective has value $f_i$ in sector $i$, it gives the claimed joint distribution with $X_{\rm out}$.

Only the reduced state of the candidate enters the calculation. Sampling garbage and other unused auxiliary registers may therefore be discarded.
\end{proof}

Only the candidate and its classical score need be retained after a trial. Sampling garbage and unused work may be discarded, by Lemma~\ref{lem:latent-sector}. Only the current candidate and the candidate produced by a new trial need coexist. The check $y<\tau$ is automatic for ideal active trials but makes the decreasing-threshold rule explicit.

\subsection{Selection guarantee}

\begin{theorem}\label{thm:seed}
Under the conditions of Section~\ref{sec:controller}, with probability at least $1-\sigma$ the algorithm returns a pair $(i,y)$ satisfying jointly
\begin{equation}
 q_i^-(y),q_i^+(y)>\gamma,\qquad y\le a_p+\eps.
 \label{eq:strong-seed}
\end{equation}
For a retained candidate this is a statement about its latent final sector, as in Lemma~\ref{lem:latent-sector}. Sufficient worst-case budgets are
\begin{equation}
 Q_P^{\rm seed}=\sum_{j=1}^K t_j3^{j-1},\qquad
 Q_A^{\rm seed}=\sum_{j=1}^K t_jq_j^A+nT,
 \label{eq:finite-seed-cost}
\end{equation}
where $q_j^A$ satisfies~\eqref{eq:trial-cost}. The live quantum space is $O(s_P+w+\log(K+1))$.
\end{theorem}

The following estimate applies to unnormalized strict ranks, including atoms, ties, and adaptively accepted proposals. It does not assume independence between successive thresholds.

\begin{lemma}[Rank-cap hitting]\label{lem:hitting}
Let $\omega$ be nonnegative weights on a finite set of acceptable pairs and $R(\tau)=\sum_{z:y<\tau}\omega(z)$. Suppose each trial at the current threshold proposes each acceptable $z$ with probability $c(\tau)^2\omega(z)\one_{\{y<\tau\}}$. Whenever the desired cap has not been reached, suppose $c(\tau)^2\ge C>0$, uniformly conditional on earlier history. An acceptable proposal is accepted with conditional probability at least $v_{\rm acc}>0$, and every accepted proposal decreases $\tau$ to its score. Other accepted proposals are allowed, but must also decrease the threshold. For $\theta>0$ with $v_{\rm acc}C\theta\le1$, after $t$ trials,
\[
 \Pr\{R(\tau)>\theta\}\le(1-v_{\rm acc}C\theta)^t
 \le\exp(-v_{\rm acc}C\theta t).
\]
\end{lemma}
\begin{proof}
If total acceptable weight is at most $\theta$, the cap already holds. Otherwise let $y_*$ be the smallest score such that acceptable pairs with $y\le y_*$ have total weight at least $\theta$. Acceptable pairs with $y<y_*$ have total weight less than $\theta$. Whenever $R(\tau)>\theta$, one has $y_*<\tau$. Thus acceptable pairs with $y\le y_*$ remain available, and accepting any of them makes $R(\tau)\le\theta$. The conditional probability of accepting one of these pairs is at least $v_{\rm acc}C\theta$. A decreasing threshold cannot undo the cap. Let $\mathcal F_r$ be the history after $r$ trials. On every history with $R(\tau_r)>\theta$, the conditional probability of missing the cap in trial $r+1$ is at most $1-v_{\rm acc}C\theta$. Iterating conditional expectation proves the first bound.
\end{proof}

\begin{proof}[Proof of Theorem~\ref{thm:seed}]
We show that after level $j$,
\begin{equation}
 R_j(\tau)\le h_j^\star9^{-(j-1)}.
 \label{eq:rank-cap}
\end{equation}
Fix any history up to the start of level $j$ for which the bounds at earlier levels hold. Since accepted scores decrease $\tau$, these bounds continue to hold during level $j$. For the next trial at this level, we therefore have
\[
 s_\ell(\tau)
 =c_\ell(\tau)^2\bigl(R_\ell(\tau)+H_\ell(\tau)\bigr)
 \le h_\ell^\star+\Lambda_\ell,
 \qquad 1\le\ell<j.
\]
By the definition of $a_j^\star$ and positivity of the factors,
\[
 \frac{c_j(\tau)}{3^{j-1}}
 =\prod_{\ell<j}\left(1-\frac43s_\ell(\tau)\right)
 \ge a_j^\star>0.
\]
Apply Lemma~\ref{lem:hitting} with
\[
 C=\frac{\underline\eta_j9^{j-1}}
          {v_{\rm acc}h_j^\star},
 \qquad
 \theta=h_j^\star9^{-(j-1)},
\]
for which $c_j(\tau)^2\ge C$ and $v_{\rm acc}C\theta=\underline\eta_j<1$. The probability that~\eqref{eq:rank-cap} first fails at level $j$ is therefore at most $(1-\underline\eta_j)^{t_j}$. Summing over $j$ gives
\[
 \Pr\{\text{at least one rank bound fails}\}
 \le\sum_{j=1}^K(1-\underline\eta_j)^{t_j}
 \le\varepsilon_{\rm cap}.
\]

Suppose all the rank bounds hold. If the final score satisfies $\tau>a_p+\eps$, including $\tau=+\infty$, Lemma~\ref{lem:mass} and~\eqref{eq:retention} give
\[
 R_K(\tau)\ge r_*(K)(1-2\beta)p
 >h_K^\star9^{-(K-1)},
\]
where the strict inequality follows from \eqref{eq:finite-certificate}. This contradicts \eqref{eq:rank-cap} at level $K$. Hence an incumbent exists and its score is at most $a_p+\eps$.

At each validation, conditional on the preceding history, the probability of accepting a pair with $\min\{q_i^-(y),q_i^+(y)\}\le\gamma$ is at most $\zeta$. By Lemma~\ref{lem:latent-sector} this bound also holds for retained quantum candidates. If such a pair is not accepted during any validation, the final incumbent must satisfy $q_i^-(y),q_i^+(y)>\gamma$. Since there are at most $T$ validations,
\[
 \Pr\{\text{\eqref{eq:strong-seed} fails}\}
 \le \varepsilon_{\rm cap}+T\zeta
 =\varepsilon_{\rm cap}+\varepsilon_{\rm val}
 =\sigma.
\]
At level $j$, a trial uses $3^{j-1}$ preparation queries and at most $q_j^A$ estimator queries, followed by at most $n$ estimator queries for validation. Summing the costs of the $t_j$ trials at each level therefore gives \eqref{eq:finite-seed-cost}. Trials and validations share the estimator workspace, so the space bound is the same as one trial plus at most $s_P$ qubits for the incumbent.
\end{proof}

\subsection{Geometric scheduling}
\label{sec:geometric-schedule}
Assume the filter certificates satisfy~\eqref{eq:leakage} with $r_*>0$ and $\Lambda<3/4$, and $d_\ell=O(\ell+1)$. Choose fixed constants
\begin{equation}
 \frac13<\rho<1,\quad 0<\vartheta<1,\quad h>0,\quad
 0<m<r_*(1-2\beta),\quad
 \frac{h}{1-\rho}+\Lambda<\frac34.
 \label{eq:geometric-conditions}
\end{equation}
Each level makes three calls to the preceding level, which accounts for the lower bound on $\rho$. Fix the validation and failure constants from Section~\ref{sec:controller} and set
\begin{equation}
 K=1+\max\left\{0,\left\lceil\log_9\frac{h}{mp}\right\rceil\right\}.
 \label{eq:depth}
\end{equation}
For distance $d\ge1$ from the last level, put
\begin{align}
 h_d&=h\rho^{d-1},&
 \delta_d&=\varepsilon_{\rm cap}(1-\vartheta)\vartheta^{d-1},
 \label{eq:cap-schedule}\\
 a_d&=1-\frac43\left(\frac{h\rho^d}{1-\rho}+\Lambda\right),&
 \eta_d&=v_{\rm acc} a_d^2 h_d.
 \label{eq:gain-schedule}
\end{align}
Choose $L_d\ge\log(1/\delta_d)$ with $L_d=\Theta(d)$, and set
\begin{equation}
 T_d=\left\lceil L_d/\eta_d\right\rceil,\quad
 t_j=T_{K-j+1},\quad h_j^\star=h_{K-j+1},\quad
 T=\sum_{d=1}^K T_d.
 \label{eq:trial-schedule}
\end{equation}
Use symmetric validation with $\widehat\zeta=\min\{\varepsilon_{\rm val}/T,1-v_{\rm acc}\}$, or any certified validation procedure with the same completeness and soundness guarantees.

\begin{corollary}\label{cor:geometric-seed}
The geometric schedule satisfies Theorem~\ref{thm:seed} and has
\begin{equation}
 Q_P^{\rm seed}=O(p^{-1/2}),\qquad
 Q_A^{\rm seed}=O(p^{-1/2}),\qquad
 S_Q^{\rm seed}=O(s_P+w+\log\log(4/p)).
 \label{eq:seed-space}
\end{equation}
The implied constants are independent of $p$, $\delta$, and the number of possible estimator outputs.
\end{corollary}
\begin{proof}
For a level $j$ at distance $d=K-j+1$, the cap and leakage estimates give
\[
 \sum_{\ell<j}(h_\ell^\star+\Lambda_\ell)
 \le\frac{h\rho^d}{1-\rho}+\Lambda<\frac34.
\]
Every product factor is positive. The elementary product bound yields $a_j^\star\ge a_d>0$, so $\eta_d\le\eta_j^\star$. Also $h<(3/4)(1-\rho)<1/2$, hence $0<\eta_d<1$. Taking $\underline\eta_j=\eta_d$ gives $(1-\underline\eta_j)^{t_j}\le e^{-\eta_dT_d}\le\delta_d$ and $\sum_{d=1}^K\delta_d\le\varepsilon_{\rm cap}$. The depth choice ensures
\[
 h9^{-(K-1)}\le mp<r_*(1-2\beta)p\le r_*(K)(1-2\beta)p,
\]
which is the final-cap condition.

The $a_d$ have a positive constant lower bound, and $T_d=O(d\rho^{-(d-1)})$. Since $3\rho>1$,
\[
 Q_P^{\rm seed}
 =3^{K-1}\sum_{d=1}^K\frac{T_d}{3^{d-1}}
 =O(3^{K-1})=O(p^{-1/2}).
\]
The degree condition and~\eqref{eq:trial-cost} give the same bound for base-estimator queries used by the coherent trials. Validation uses $O(T\log(2T))$ forward base-estimator queries. Since $T=O(K\rho^{-K})$ and $K=\log_9(1/p)+O(1)$ as $p\to0$, this is
\[
 O\!\left(p^{-\nu}\log^2(2/p)\right)=o(p^{-1/2}),\qquad
 \nu=\frac{\log(1/\rho)}{\log9}<\frac12.
\]
Finally, $\log(K+1)=O(\log\log(4/p))$.
\end{proof}

For every fixed admissible margin triple, Appendix~\ref{app:filter-feasibility} provides geometric-error filters with $r_*>0$ and arbitrarily small $\Lambda$. Thus the schedule conditions are feasible. Fixing any one universal admissible choice makes the hidden constants universal in Theorem~\ref{thm:main}. Appendix~\ref{app:concrete-parameters} fixes one admissible filter family and selection parameters; Appendix~\ref{app:concrete-schedule} derives finite schedules directly from Theorem~\ref{thm:seed} for each problem size.

\subsection{Confidence amplification}\label{sec:confidence}

On the success event of Theorem~\ref{thm:seed}, the returned pair satisfies $q_i^-(y),q_i^+(y)>\gamma>1/3$. This margin above $1/3$ allows confidence amplification with outer validation while retaining a valid candidate. Fix a choice of the constant parameters in Corollary~\ref{cor:geometric-seed}, with failure bound $\sigma<1/2$. Choose positive constants $\omega_{\rm bad},\omega_{\rm none}, \omega_{\rm rej}$ whose sum is at most one. For $\delta<\sigma$, set
\begin{equation}
 R=\left\lceil\frac{\log(1/(\omega_{\rm none}\delta))}
                         {\log(1/\sigma)}\right\rceil,\qquad
 \xi=\min\left\{\frac{\omega_{\rm bad}\delta}{R},
                      \omega_{\rm rej}\delta\right\}.
 \label{eq:outer-schedule}
\end{equation}
Then $R\xi+\sigma^R+\xi\le\delta$. The confidence-amplification procedure uses $p,\delta$ and fixed design parameters; $\eps$ is used only to form the final interval.

\begin{algorithm}
\label{alg:confidence}
\leavevmode
\begin{enumerate}[label=\arabic*.,leftmargin=2em,itemsep=3pt]
\item Use the geometric schedule from Corollary~\ref{cor:geometric-seed} for Algorithm~\ref{alg:seed}, with failure bound $\sigma$ from Theorem~\ref{thm:seed}. If $\delta\ge\sigma$, run the seed algorithm once, store its output in $\mathsf{best}$, and go to Step~4.

\item For $\delta<\sigma$, compute $R$ and $\xi$ from~\eqref{eq:outer-schedule} and set $\mathsf{best}\leftarrow\bot$.

\item Repeat the following steps $R$ times
\begin{enumerate}[label=(\alph*),leftmargin=1.8em,itemsep=2pt]
\item Run Algorithm~\ref{alg:seed} with the schedule chosen in Step~1. If it returns $\bot$, continue to the next iteration.
\item Validate the returned candidate and score $y$ using $\Val_{1/3,\gamma}(y,\xi)$ from Lemma~\ref{lem:validation}, with $n_{1/3,\gamma}(\xi)$ new estimates as specified in~\eqref{eq:validation-count}. If the validation procedure rejects the score $y$, discard the candidate and continue to the next iteration.
\item Use the same incumbent rule as in Algorithm~\ref{alg:seed}. If $\mathsf{best}=\bot$ or $y$ is smaller than the best score, replace $\mathsf{best}$ with the new candidate and score. Otherwise, discard the new candidate.
\end{enumerate}

\item If $\mathsf{best}=\bot$, prepare a new candidate using the unitary $P$ from Section~\ref{sec:model} and set $v=0$. Otherwise, take the candidate and score $v$ from $\mathsf{best}$. Return the candidate and
\[
 (v,L,U)=(v,v-\eps,v+\eps),
\]
in the form specified by Theorem~\ref{thm:main}.
\end{enumerate}
\end{algorithm}

\begin{proof}[Proof of Theorem~\ref{thm:main}]
By Theorem~\ref{thm:seed}, an output satisfying \eqref{eq:strong-seed} is obtained with probability at least $1-\sigma$. Since $\gamma>1/3$, the inequalities for $q_i^-(y)$ and $q_i^+(y)$ imply $|y-f_i|\le\eps$. Together with $y\le a_p+\eps$, this gives \eqref{eq:main-success}. In case $\delta\ge\sigma$, the required probability is therefore obtained by running Algorithm~\ref{alg:seed} once.

Assume now that $\delta<\sigma$. By Lemma~\ref{lem:latent-sector}, we can compute the validation probabilities conditional on the sector of each candidate. If $|y-f_i|>\eps$, then at least one of $q_i^-(y)$ and $q_i^+(y)$ is at most $1/3$, and Lemma~\ref{lem:validation} gives probability at most $\xi$ of accepting the score. The probability of accepting an inaccurate score in any of the $R$ runs is therefore at most $R\xi$. Since the runs are independent, the probability that none of the outputs satisfies~\eqref{eq:strong-seed} is at most $\sigma^R$. If one or more outputs satisfy these inequalities, consider the first of them. Conditional on the seed outputs and their sectors, Lemma~\ref{lem:validation} gives probability at most $\xi$ of rejecting this output. It follows that, with probability at least $1-R\xi-\sigma^R-\xi\ge1-\delta$, an output with score at most $a_p+\eps$ is accepted and all accepted scores are accurate. Selecting the smallest accepted score then gives \eqref{eq:main-success}.

Let $B_P,B_A=O(p^{-1/2})$ be the query bounds for one run of Algorithm~\ref{alg:seed}, as given by Corollary~\ref{cor:geometric-seed}. For $\delta<\sigma$, the choices in~\eqref{eq:outer-schedule} give $R=O(u)$ and $\log(1/\xi)=O(u)$, where $u=\log(2/\delta)$. Since each validation uses $O(\log(1/\xi))$ estimator calls, we obtain
\[
 \begin{aligned}
 Q_P&\le RB_P+1=O(p^{-1/2}u),\\
 Q_A&\le R\bigl(B_A+O(\log(1/\xi))\bigr)
       =O(p^{-1/2}u+u^2).
 \end{aligned}
\]
By Corollary~\ref{cor:geometric-seed}, Algorithm~\ref{alg:seed} uses $S_Q^{\rm seed}=O(s_P+w+\log\log(4/p))$ qubits. These registers can be reused in each repetition, and validation uses the estimator work registers. The best candidate from earlier repetitions is stored separately in at most $s_P$ qubits. The numbers of repetitions and estimates collected so far are stored classically. The total number of qubits therefore satisfies
\[
 S_Q\le S_Q^{\rm seed}+s_P
 =O(s_P+w+\log\log(4/p)).
\]
\end{proof}

\section{Applications}
\label{sec:applications}

The theorem can be used with any estimator satisfying Definition~\ref{def:oracle}. If each application of $A$ or $A^\dagger$ requires at most $q(\eps)$ queries to an underlying oracle, substituting this cost into Theorem~\ref{thm:main} gives
\begin{equation}
 Q_P=O(p^{-1/2}u),\qquad
 Q_{\rm prim}=O\bigl(q(\eps)[p^{-1/2}u+u^2]\bigr),
 \qquad u=\log(2/\delta).
 \label{eq:primitive}
\end{equation}
We apply this substitution to amplitude estimation for bounded expectations and phase estimation for energy selection. Neither application requires controlled calls to the estimator or its inverse.

\subsection{Bounded-expectation selection}
\label{sec:expectations}
Let $W=\sum_{i=1}^N\ket i\bra i\otimes W_i$ be a label-preserving unitary with $W,W^\dagger$ available, and let $\Pi_1$ be a known projector on its work register. The objective is
\begin{equation}
 f(i)=\norm{(I\otimes\Pi_1)W\ket i\ket0}^2\in[0,1].
 \label{eq:expectation}
\end{equation}

For example, consider a reversible simulation with candidate label $i$ as input. Measuring its loss register would give a random value $Z_i\in[0,1]$. For each possible value $z$ of $Z_i$, a rotation controlled by the loss register maps an auxiliary qubit from $\ket0$ to $\sqrt{1-z}\ket0+\sqrt z\ket1$. Taking $\Pi_1=\ket1\bra1$ on this qubit gives $f(i)=\mathbb E[Z_i]$ in~\eqref{eq:expectation}.

By using amplitude estimation~\cite{BHMT}, we obtain a coherent estimator that is $\eps$-accurate with probability at least $2/3$, using $O(1/\eps)$ calls to $W$ and $W^\dagger$. Both $W$ and $W^\dagger$ preserve the label, and the remaining operations act only on the work register. To control a Grover operator, we place controls on the known reflections, leaving the calls to $W$ and $W^\dagger$ uncontrolled. When the control qubit is $0$, the reflections act as the identity and the two oracle calls cancel.

Standard phase-estimation-based amplitude estimation uses $O(w_W+\log(2/\eps))$ work qubits, where $w_W$ is the work width of $W$. Classical samples alone do not provide coherent access to $W$ and $W^\dagger$.

\begin{corollary}
\label{cor:prior-expectation}
Assume~\eqref{eq:expectation}, and let $P$ independently prepare a prior $\mu$ on the labels. For known $0<p\le1$, $0<\eta<1$, and $0<\delta<1/2$, an algorithm returns a label $s$ such that, simultaneously for every $a$ with $\mu\{f\le a\}\ge p$,
\[
 \Pr\{f(s)\le a+\eta\}\ge1-\delta.
\]
It also returns the certificate~\eqref{eq:certificate} at numerical precision $\eps=\eta/2$. With $u=\log(2/\delta)$, its worst-case query bounds are
\begin{equation}
 Q_P=O(p^{-1/2}u),\qquad
 Q_W=O\bigl(\eta^{-1}(p^{-1/2}u+u^2)\bigr).
 \label{eq:expectation-cost}
\end{equation}
Its live quantum space is $O(s_P+w_W+\log(2/\eta)+\log\log(4/p))$. Neither controlled $P$ nor controlled $W$ is required.
\end{corollary}
\begin{proof}
Use amplitude estimation at precision $\eps=\eta/2$ as the base estimator in Theorem~\ref{thm:main}. Its exact label-preserving form satisfies Definition~\ref{def:oracle}. Substitute $q(\eps)=O(1/\eta)$ in~\eqref{eq:primitive}. The space bound follows from the estimator width. The successful event of the main theorem implies all the stated target inequalities jointly.
\end{proof}

For uniform preparation, the constant-confidence query bound is worst-case optimal.

\begin{corollary}
\label{thm:expectations}
For $N\ge2$ and $0<\eta<1$, returning an $\eta$-optimal label,
\[
 f(s)\le\min_i f(i)+\eta,
\]
with probability at least $2/3$ has worst-case query complexity $\Theta(\sqrt N/\eta)$ in the model~\eqref{eq:expectation}. Failure at most $\delta$ is achieved with
\[
 Q_W=O\bigl(\eta^{-1}(\sqrt N\,u+u^2)\bigr),
 \qquad u=\log(2/\delta),
\]
on every execution and $O(w_W+\log(2N)+\log(2/\eta))$ live qubits. The bounds also hold with controlled access to $W$ and $W^\dagger$.
\end{corollary}
\begin{proof}
With uniform preparation, each label has probability $1/N$, so the total probability assigned to minimizing labels is at least $1/N$. Applying Corollary~\ref{cor:prior-expectation} with $p=1/N$ therefore gives a label $s$ satisfying $f(s)\le\min_i f(i)+\eta$ with probability at least $1-\delta$. Uniform preparation uses $O(\log(2N))$ qubits and requires no calls to $W$ or $W^\dagger$. Since $p=1/N$, the additional term $\log\log(4/p)=\log\log(4N)$ is also $O(\log(2N))$. The matching lower bound is proved in Appendix~\ref{app:expectation-lower}.
\end{proof}
For $\eta\ge1$, any label is $\eta$-optimal since $f(i)\in[0,1]$. The corresponding maximization result follows by replacing $\Pi_1$ with $I-\Pi_1$.

Table~\ref{tab:expectation-comparison} compares query complexity of the proposed method with existing methods for selecting an $\eta$-optimal label. For the algorithm of Wang, You, Li, and Childs~\cite[Corollary 1]{WYLC}, the query bound also depends on the differences between the minimum objective value and the remaining values. Their bound assumes a unique minimum. Ordering the losses as $f_{(1)}<f_{(2)}\le\cdots\le f_{(N)}$, define
\[
 \Delta_i=f_{(i)}-f_{(1)},\qquad
 H=\sum_{i=2}^N\Delta_i^{-2}.
\]
For fixed positive gaps, their bound becomes independent of $\eta$ when $\eta\le\Delta_2$, since the minimum of $N/\eta^2$ and $H$ is then $H$. Corollary~\ref{thm:expectations} also applies when several labels attain the minimum.

\begin{table}[tb]
\centering\small
\caption{Query and sample bounds for selecting an $\eta$-optimal label with fixed success probability, for $N\ge2$ and $0<\eta<1$. The classical bound counts independent samples; the quantum bounds count calls to the supplied coherent oracle and its inverse.}
\label{tab:expectation-comparison}
\begin{tabular}{@{}>{\raggedright\arraybackslash}p{.46\textwidth}>{\raggedright\arraybackslash}p{.48\textwidth}@{}}
\toprule
Method & Queries or samples\\
\midrule
Classical median elimination~\cite{EMM} & $O(N/\eta^2)$\\
\addlinespace
Confidence amplification before minimum finding~\cite{BHMT,CDW} & $O((\sqrt N/\eta)\log(2N))$\\
\addlinespace
Approximate quantum best-arm identification~\cite{WYLC} & $O\!\left(\sqrt{\min\{N/\eta^2,H\}}\,\polylog(N/\Delta_2)\right)$\\
\addlinespace
SOGAS~\cite{HHZ} & $O((\sqrt N/\eta)\polylog(N/\eta))$\\
\addlinespace
Corollary~\ref{thm:expectations} & $\Theta(\sqrt N/\eta)$\\
\bottomrule
\end{tabular}
\end{table}

The classical bound follows from median elimination~\cite[Theorem 10]{EMM} applied to rewards $1-Z_i$. For quantum minimum finding, using amplitude estimation with $\eps=\eta/2$ in Proposition~\ref{prop:amplified-general}, with $p=1/N$, results in $O((\sqrt N/\eta)\log(2N))$ queries. The factor $\log(2N)$ comes from repeated calls to the estimator to reduce its failure probability.

The analysis of SOGAS by Hu, Hu, and Zhou~\cite[proof of Theorem 1 in Appendix G]{HHZ} gives $O((\sqrt N/\eta)\polylog(N/(\delta\eta)))$ queries for coherent simulation with bounded output. For fixed $\delta$, this upper bound has the same $\sqrt N/\eta$ factor as Corollary~\ref{thm:expectations}, together with additional factors polynomial in $\log(N/\eta)$.

\subsection{Low-energy state preparation}
\label{sec:spectral}
Let $0\le H\le I$, with ground energy $E_0=\lambda_{\min}(H)$, and let $P$ prepare a guiding state, possibly entangled with reference registers. We assume access to $P,P^\dagger$ and controlled applications of $\mathcal U=e^{iH}$ and its inverse. Each controlled application counts as one evolution query, so implementing a controlled power $\mathcal U^k$ requires $|k|$ queries.

For $0<\eps<1$, coherent phase estimation~\cite{Kitaev,BHMT} estimates the energy to accuracy $\eps$ with probability at least $2/3$, using $O(1/\eps)$ evolution queries and $O(\log(2/\eps))$ work qubits. We decode the phase estimates in $[-\pi,\pi)$, which contains the energy interval $[0,1]$ in its interior. Estimates outside of this energy interval are clipped to $0$ or $1$. This replacement does not increase the estimation error, because the true energy is in $[0,1]$. On an eigenspace with energy $E$, the evolution $\mathcal U$ acts as $e^{iE}I$, so the controlled evolutions act on the estimator's work register in the same way for every state in that eigenspace. The estimator therefore has the form
\[
 A=\sum_i\Pi_i\otimes A_i,
\]
where $\Pi_i$ are the energy projectors, and satisfies Definition~\ref{def:oracle} also when the eigenvalues are degenerate. Its construction uses only the evolution oracle and known gates.

\begin{corollary}
\label{cor:spectral}
Suppose the guiding state has probability at least $p$ on energies at most $a$, where $0<p\le1$ is known. For $0<\eta<1$ and $0<\delta<1/2$, an algorithm returns a candidate whose density matrix, averaged over the algorithm's measurement outcomes and classical random choices, has weight at least $1-\delta$ on energies at most $a+\eta$. It also returns the certificate~\eqref{eq:certificate} with $\eps=\eta/2$. The query bounds are
\begin{equation}
 Q_P=O(p^{-1/2}u),\qquad
 Q_{\mathcal U}
 =O\bigl(\eta^{-1}(p^{-1/2}u+u^2)\bigr),
 \qquad u=\log(2/\delta).
 \label{eq:spectral-cost}
\end{equation}
The quantum space is $O(s_P+\log(2/\eta)+\log\log(4/p))$ qubits, excluding workspace internal to the supplied evolution oracle. The threshold $a$ and the spectral distribution need not be known.
\end{corollary}

\begin{proof}
Apply Theorem~\ref{thm:main} with the phase estimator above and $\eps=\eta/2$. Since the guiding state has probability at least $p$ on energies at most $a$, we have $a_p\le a$. It follows that
\[
 X_{\rm out}\le a_p+2\eps\le a+\eta
\]
with probability at least $1-\delta$, which gives the stated weight in the averaged output density matrix. The estimator and its inverse each require $O(1/\eta)$ evolution queries and $O(\log(2/\eta))$ work qubits.  The query and space bounds follow by substituting these costs into Theorem~\ref{thm:main}.
\end{proof}

Suppose now that the guiding state has ground-space amplitude overlap at least $\omega>0$, and that a positive lower bound $\Delta$ on the spectral gap is known. Choose a fixed $0<c_{\rm gap}<1/2$ and apply Theorem~\ref{thm:main} with $p=\omega^2$ and $\eps=c_{\rm gap}\Delta$. Since the probability in the ground space is at least $\omega^2$, we have $a_p=E_0$. The theorem therefore gives
\[
 X_{\rm out}\le E_0+2c_{\rm gap}\Delta<E_0+\Delta
\]
with probability at least $1-\delta$. Every energy outside the ground space is at least $E_0+\Delta$, so the averaged output density matrix has ground-space weight at least $1-\delta$. The query bounds become
\[
 Q_P=O(\omega^{-1}u),\qquad
 Q_{\mathcal U}
 =O\bigl(\Delta^{-1}(\omega^{-1}u+u^2)\bigr).
\]
Ground-state preparation from a guiding state has also been studied by Jeffery, Somma, Witteveen, and de Wolf~\cite{JSWW}. For a Hamiltonian with a unique ground state, they give a transducer construction and an algorithm combining amplitude amplification with verification~\cite[Theorem~1.3 and Section~2]{JSWW}. Both use an estimate of the ground energy, which can first be obtained using the algorithm of Jeffery and Witteveen~\cite[Theorem~1.1]{JW}. Chen, Gao, Wang, and Zhou~\cite{CGWZ} study ground-state preparation using a Hamiltonian block encoding, assuming a unique ground state and a known energy threshold inside the gap. Both works bound the trace distance between the prepared state and the target state. The result above gives $\Pr\{X_{\rm out}=E_0\}\ge1-\delta$ and applies to ground spaces of any dimension.

\subsection{Ground-energy estimation}
\label{sec:energy}

Consider the same Hamiltonian access model, but with the goal of only obtaining a classical energy estimate. Repeating the minimum-finding algorithm of Theorem~\ref{thm:main} independently and taking the median of the resulting estimates reduces the dependence on $\log(2/\delta)$ in the evolution-query bound from quadratic to linear.

\begin{corollary}
\label{cor:energy}
Suppose $\norm{(\Pi_{\min}\otimes I_R)P\ket0}\ge\omega>0$, where $\Pi_{\min}$ is the ground-space projector. Given this lower bound, $0<\eta<1/4$, and $0<\delta<1/2$, an additive-$\eta$ estimate of $E_0$ can be returned with probability at least $1-\delta$ using
\begin{equation}
 Q_P=O(\omega^{-1}u),\qquad
 Q_{\mathcal U}=O((\omega\eta)^{-1}u),
 \qquad u=\log(2/\delta),
 \label{eq:energy-cost}
\end{equation}
on every execution, with no requirement of a spectral gap, initial energy estimate, or controlled preparation. The algorithm uses $O(s_P+\log(2/\eta)+\log\log(4/\omega^2))$ qubits, with the estimates and repetition counter stored classically. As in Corollary~\ref{cor:spectral}, the qubit count excludes the internal workspace of the evolution oracle.
\end{corollary}
\begin{proof}
Set $p=\omega^2$. Since the ground-space mass is at least $p$, the quantile in~\eqref{eq:quantile} is $a_p=E_0$. Fix $0<\sigma_E<1/2$ and run Theorem~\ref{thm:main} at failure probability $\sigma_E$ with phase-estimation precision $\eps=\eta$. On its successful event,
\[
 E_0-\eta\le X_{\rm out}-\eta\le v\le a_p+\eta=E_0+\eta.
\]
Thus with probability at least $1-\sigma_E$, the marginal classical estimate $v$ is within $\eta$ of $E_0$. Each run uses $O(\omega^{-1})$ preparation queries and $O((\omega\eta)^{-1})$ evolution queries. Repeat the algorithm independently $m$ times, where $m$ is an odd number, to obtain estimates $v_1,\ldots,v_m$, and return their median. If more than half of these estimates are in $[E_0-\eta,E_0+\eta]$, then their median belongs to the same interval. Since each run fails with probability at most $\sigma_E<1/2$, a Chernoff bound gives
\[
 \Pr\{|\operatorname{median}(v_1,\ldots,v_m)-E_0|>\eta\}
 \le \exp\!\left(-2m(1/2-\sigma_E)^2\right).
\]
By choosing $m=O(u)$, the right-hand side is at most $\delta$, which gives~\eqref{eq:energy-cost}. The quantum registers can be reused after recording each estimate.
\end{proof}

Table~\ref{tab:energy-comparison} compares the query bounds for a fixed success probability. Jeffery and Witteveen~\cite[Theorem 1.1]{JW} obtain $O((\omega\eta)^{-1})$ controlled-evolution queries. Corollary~\ref{cor:energy} gives the same evolution bound with $O(\omega^{-1})$ calls to $P$ and $P^\dagger$, improving the preparation upper bound in their theorem. To compare the evolution queries, their phase convention can be converted to ours by a constant phase shift and reversal of the evolution. Their theorem assumes controlled preparation and also gives elementary-gate and workspace bounds.

\begin{table}[tb]
\centering\small
\caption{Upper bounds for guided ground-energy estimation with fixed success probability, precision $\eta$, and ground-space amplitude overlap at least $\omega$. Here $\ell_\eta=\log(2/\eta)$. Each elementary controlled-evolution operation counts as one evolution query, and both query counts include inverse applications. The AGGW bounds follow from the phase-estimation construction described below.}
\label{tab:energy-comparison}
\begin{tabular}{@{}>{\raggedright\arraybackslash}p{.39\textwidth}>{\raggedright\arraybackslash}p{.28\textwidth}>{\raggedright\arraybackslash}p{.25\textwidth}@{}}
\toprule
Construction & Preparation queries & Evolution queries\\
\midrule
AGGW with upfront-amplified phase estimation & $O(\omega^{-1})$ & $O(\log(2/\omega)/(\omega\eta))$\\
\addlinespace
Jeffery--Witteveen~\cite{JW} & $O(\omega^{-1}\ell_\eta\log(2\ell_\eta))$ & $O((\omega\eta)^{-1})$\\
\addlinespace
Corollary~\ref{cor:energy} & $O(\omega^{-1})$ & $O((\omega\eta)^{-1})$\\
\bottomrule
\end{tabular}
\end{table}

For the first row, use Proposition~\ref{prop:amplified-general} with $p=\omega^2$ and a phase estimator of precision $\eta$. Its accurate low estimate is within $\eta$ of $E_0$ on the successful event, by the same inequality used above. Each call to the confidence-amplified phase estimator uses $O(\eta^{-1}\log(2/\omega))$ elementary controlled-evolution queries. This comparison uses the controlled-evolution model, whereas AGGW's published energy application~\cite[Lemma 50]{AGGW} is stated for sparse-Hamiltonian access.

Mande and de Wolf~\cite{MDW} establish phase-estimation and related lower bounds. Somma and de Wolf~\cite[Theorem 3.1]{SDW} construct hard Hamiltonian families of dimension $O(\omega^{-2}\log(1/\delta))$ requiring $\Omega((\omega\eta)^{-1}\log(1/\delta))$ controlled-evolution queries, even with unlimited preparation access. After constant rescaling into $[0,1]$, this gives the matching worst-case lower bound for~\eqref{eq:energy-cost} when $0<\omega,\eta,\delta\le c$ and the allowed dimension is at least $C\omega^{-2}\log(1/\delta)$, for suitable universal constants $c>0$ and $C>0$. Thus, the evolution-query complexity is optimal in its dependence on overlap, precision, and confidence within the stated parameter range.

\section{Conclusion}\label{sec:discussion}

This paper has developed a quantum minimum-finding algorithm for objective values accessed through coherent numerical estimators that avoids the logarithmic overhead of amplifying their confidence before the search. It applies to arbitrary prepared distributions and allows the selected candidate to be retained as a quantum state. Assume the preparation assigns probability mass at least a known $p$ to objective values at most $a$. The algorithm uses $O(p^{-1/2}\log(2/\delta))$ preparation queries and $O(p^{-1/2}\log(2/\delta)+\log^2(2/\delta))$ estimator queries to select a candidate with objective value at most $a+2\eps$ with probability at least $1-\delta$. At constant success probability, the estimator-query bound and the factor $2$ in the selection error are separately worst-case optimal.

Uniform preparation over $N$ classical labels gives $p=1/N$ and hence $O(\sqrt N)$ estimator queries with additive selection error $2\eps$. Combining this result with amplitude estimation gives a worst-case optimal complexity of $\Theta(\sqrt N/\eta)$ queries to the underlying oracle for selecting an $\eta$-optimal bounded expectation.

Using phase estimation, a candidate with energy at most $a+\eta$ can be selected with $O(p^{-1/2})$ preparation queries and $O(p^{-1/2}/\eta)$ elementary controlled-evolution queries. A guiding state with ground-space amplitude overlap at least $\omega$ gives ground-energy estimation to additive error $\eta$ using $O(\omega^{-1})$ preparation queries and $O((\omega\eta)^{-1})$ elementary controlled-evolution queries.

Further work could reduce the constants by improving the polynomial filters and validation parameters. It also remains open whether the $\log^2(2/\delta)$ term can be removed from the estimator-query bound when the selected quantum candidate is retained.

\section*{Author contribution statement}

The contributions presented in this paper were developed by the author. OpenAI's ChatGPT was used to assist with surveying the literature, discussing mathematical ideas, and editing the manuscript.

\section*{Acknowledgements}

This work was supported in part by the Research Council of Norway through the project Quantum computing for the defence sector -- QC4DS (grant no. 346299).

\newpage
\appendix
\section{Confidence amplification before minimum finding}
\label{app:upfront-comparison}

This appendix analyzes implementations of AGGW~\cite{AGGW} and Chen--de Wolf~\cite{CDW} that amplify the estimator's confidence before minimum finding, under the oracle conventions in Definition~\ref{def:oracle}. The resulting query bounds are given in Table~\ref{tab:minimum-comparison}. We first consider generalized minimum finding with arbitrary preparation, and then the specialization to uniform preparation over labels.

\begin{proposition}
\label{prop:amplified-general}
Suppose $\mu\{f\le a\}\ge p$. Using upfront confidence amplification, AGGW generalized minimum finding~\cite[Algorithm 3 and Theorem 49]{AGGW} can return a retained candidate and a numerical value $y$ such that, jointly with probability at least $2/3$,
\[
 |y-X_{\rm out}|\le\eps,\qquad y\le a+\eps,
 \qquad X_{\rm out}\le a+2\eps.
\]
The algorithm is given $p$ but not $a$ or $\mu$. On every execution it uses
\[
 Q_P=O(p^{-1/2}),\qquad Q_A=O(p^{-1/2}\log(2/p)).
\]
It requires no controlled $P$ or controlled $A$.
\end{proposition}
\begin{proof}

For $0<b<1/2$, apply the base estimator using $k=O(\log(2/b))$ separate work registers initialized to $\ket0$, where $k$ is odd and sufficiently large. Computing the median reversibly gives a unitary estimator $A_b$ with error probability at most $b$ in each sector.

Choose $b=cp$, where $c>0$ is a sufficiently small universal constant. In sector $i$, write $\ket{\chi_i}=A_{b,i}\ket0$, let $G_i$ project onto estimates within $\eps$ of $f_i$, and define
\[
 \ket{\chi_i^g}=\frac{G_i\ket{\chi_i}}{\norm{G_i\ket{\chi_i}}}.
\]
Their inner product is real and at least $\sqrt{1-b}$. Let $R_i$ be a rotation in their span that maps $\ket{\chi_i}$ to $\ket{\chi_i^g}$, and extend it by the identity on the orthogonal complement. Then
\[
 \norm{R_i-I}
 =\sqrt{2-2\langle\chi_i^g\mid\chi_i\rangle}
 \le\sqrt{2b}.
\]
Let $R=\sum_i\Pi_i\otimes R_i$ and define actual and ideal preparations
\[
 B=A_b(P\otimes I),\qquad B'=RB.
\]
Then $\norm{B-B'}\le\sqrt{2b}$, with the same bound for the inverses. The ideal preparation is used only in the analysis, its definition need not be known to the algorithm. It preserves sector masses, and all its encoded estimates are accurate. Its score law therefore assigns mass at least $p$ to estimates at most $a+\eps$.

AGGW's capped generalized minimum-finding algorithm, with a fixed cap $M=O(p^{-1/2})$, returns such a low score with probability at least $3/4$~\cite[Theorem 49]{AGGW}. Retain the registers associated with each incumbent score rather than discarding the candidate. Completed ideal proposals are restrictions of the prepared state by predicates on that score, so they remain supported on accurate sector/score pairs. In particular, a returned low score satisfies the stated candidate guarantee.

A query hybrid for the capped computation bounds the change in the probability of any final event by at most
\[
 M\norm{B-B'}=O(p^{-1/2}\sqrt b)=O(\sqrt c).
\]
This holds for measured adaptive implementations by contractivity of trace distance, and the final hypothetical objective measurement can be included in the event. Taking $c$ sufficiently small makes the change at most $1/12$, giving success at least $2/3$ for the actual preparation.

Each call to $B$ or $B^\dagger$ uses one preparation query and $O(\log(2/b))=O(\log(2/p))$ base-estimator queries. The number of calls is capped at $M=O(p^{-1/2})$, giving the worst-case query bounds. Prepared-state reflections use $B,B^\dagger$ around known reflections and require no controlled access to the original unknown unitaries.
\end{proof}

For uniform preparation over $N$ labels, Chen and de Wolf~\cite[arXiv v2, Theorem 2.4 and Appendix A]{CDW} use $O(\sqrt N)$ calls to an estimator with failure probability at most $b/N$, where $b>0$ is a sufficiently small constant. Each call is implemented by computing the median reversibly from $O(\log(2N))$ base estimates. The total query count is therefore $O(\sqrt N\log(2N))$, with selection error $2\eps$ at fixed success probability. The corresponding numerical bound in Table~\ref{tab:all-budgets} is calculated in Appendix~\ref{app:cdw-constants}.

For $k=1$, the weak and strong approximation definitions of Gao, Ji, and Wang~\cite{GJW} coincide. Their Algorithm~1 has a single iteration, starting with an empty excluded-index set. The hiding operation is therefore the identity. The algorithm then applies generalized minimum finding to a uniform label preparation followed by the numerical oracle, with target mass $1/N$. This is the same reduction used by Chen and de Wolf~\cite[Appendix~A]{CDW}.

\section{Implementation of filters}
\label{app:filter-construction}

This appendix specifies the construction of bounded polynomial approximations to the sign function used in Section~\ref{sec:filters}, and verifies the filter conditions in Corollary~\ref{cor:geometric-seed}. The final subsection shows that approximating the known gates changes the success probability by at most the sum of their operator-norm errors.

\subsection{Bounded sign polynomials}\label{app:polynomial}
For an odd integer $n\ge3$ and an integer $r\ge1$, let $\mathcal T_n$ be the Chebyshev polynomial of the first kind and define
\begin{equation}
 K_{n,r}(x)=\left(\frac{\mathcal T_n(x)}{nx}\right)^{2r},\qquad
 \varphi_{n,r}(x)=\frac{\int_0^xK_{n,r}(t)\,dt}{\int_0^1K_{n,r}(t)\,dt},
 \qquad d_{n,r}=2r(n-1)+1.
 \label{eq:kernel-polynomial}
\end{equation}
The quotient at zero is defined by continuity. For odd $n$, $\mathcal T_n(x)=(-1)^{(n-1)/2}\sin(n\arcsin x)$. Thus the kernel is a nonnegative even polynomial with value one at zero. Its normalized integral is odd, nondecreasing, bounded by one on $[-1,1]$, and has exactly the displayed degree.

\begin{lemma}\label{lem:sign-polynomial}
For $0<g<1$, the worst sign error on $g\le|x|\le1$ is
\begin{equation}
 E_{n,r}(g)=\frac{\int_g^1K_{n,r}(t)\,dt}{\int_0^1K_{n,r}(t)\,dt}
 \le\frac{35\sqrt r}{16(2r-1)}(ng)^{-(2r-1)}.
 \label{eq:exact-sign-error}
\end{equation}
For rational $g$, the exact error is rational and computable by polynomial integration. For every $0<u<1$, choices with $n=O(1/g)$ and $r=O(\log(2/u))$ give $E_{n,r}(g)\le u$ and $d_{n,r}=O(g^{-1}\log(2/u))$.
\end{lemma}
\begin{proof}
Monotonicity and oddness show that the maximum error occurs at $x=g$ or $x=-g$, giving the integral expression. For $0\le t\le1/(2n\sqrt r)$ put $z=n\arcsin t$. Since $n\ge3$,
\[
 z^2\le\frac{n^2t^2}{1-t^2}\le\frac9{35r},\qquad
 \frac{\sin z}{nt}\ge1-\frac3{70r}.
\]
The second inequality uses $z\ge nt$, $\sin z\ge z-z^3/6$, and positivity of the displayed lower bound. The even exponent removes the sign of $\mathcal T_n$, and Bernoulli's inequality gives $K_{n,r}(t)\ge(1-3/(70r))^{2r}\ge32/35$. Hence
\[
 \int_0^1K_{n,r}(t)\,dt\ge\frac{16}{35n\sqrt r}.
\]
For $t\in[g,1]$, use $|\mathcal T_n(t)|\le1$. Extending the integral of the bound $(nt)^{-2r}$ to infinity gives
\[
 \int_g^1K_{n,r}(t)\,dt
 \le\frac{n^{-2r}g^{-(2r-1)}}{2r-1},
\]
which proves the inequality. Polynomial integration proves rationality. For the degree claim, fix any constant $c>1$ and take the least odd $n\ge\max\{3,c/g\}$. Then $ng\ge c$ and $n=O(1/g)$. The error bound decreases exponentially in $r$, so an integer $r=O(\log(2/u))$ suffices. Constants here may depend on the fixed $c$.
\end{proof}

\subsection{Filter guarantees}\label{app:filter-feasibility}
For fixed margins, take $u_\ell=u_0\chi^\ell$ with $0<\chi<1$ and $u_0>0$ sufficiently small. Choose the polynomials by Lemma~\ref{lem:sign-polynomial}, and set $\mathsf r_\ell=(1-u_\ell/2)^4$ and $\mathsf b_\ell=(u_\ell/2)^2$. If $u_0/[2(1-\chi)]<1$ and $9u_0^2/4<1$, then
\begin{align*}
 \inf_{j\ge1}r_*(j)
 &\ge\left(1-\frac{u_0}{2(1-\chi)}\right)^4>0,\\
 \sum_{j\ge1}\Lambda_j
 &\le\frac{u_0^2/4}{1-9u_0^2/4}.
\end{align*}
The first bound follows from $\prod(1-x_\ell)\ge1-\sum x_\ell$. For the second, bound every $\mathsf b_\ell$ by $u_0^2/4$ and sum the geometric series. Thus total leakage can be made arbitrarily small while retention stays positive. Moreover $d_\ell=O(\ell+1)$. This proves feasibility of the symbolic assumptions in Corollary~\ref{cor:geometric-seed} for every fixed admissible margin triple.

\subsection{Implementation precision}
\label{app:precision}
The results above assume exact known gates. We now estimate the change in the output when these gates are replaced by unitary approximations.

\begin{proposition}
\label{prop:gate-hybrid}
Suppose a finite adaptive quantum computation uses at most $M$ known unitary gates on each execution. Let $V_{t,h}$ be the gate at location $t$ after classical history $h$, and suppose that its approximation satisfies
\[
 \norm{\widetilde V_{t,h}-V_{t,h}}\le\epsilon_t
\]
for every possible $h$. Let $\rho_{\rm out}$ be the output density matrix of the original computation and $\widetilde\rho_{\rm out}$ that of the same computation with the known gates replaced by their approximations. Both density matrices include the quantum registers and the recorded classical outcomes. Then
\[
 D(\widetilde\rho_{\rm out},\rho_{\rm out})
 :=\tfrac12\norm{\widetilde\rho_{\rm out}-\rho_{\rm out}}_1
 \le\sum_{t=1}^M\epsilon_t.
\]
The same bound holds for the change in probability of any measurement outcome on these outputs.
\end{proposition}

\begin{proof}
For any density matrix $\rho$, the triangle inequality gives
\[
 \tfrac12\norm{\widetilde V\rho\widetilde V^\dagger
              -V\rho V^\dagger}_1
 \le\norm{\widetilde V-V}.
\]
By keeping the measurement outcomes in a classical register, the gate at location $t$ can be written as a direct sum over the histories $h$. The operator norm of the difference between the two direct sums is $\max_h\norm{\widetilde V_{t,h}-V_{t,h}}\le\epsilon_t$.

By adding identity gates after termination, it can be assumed that every execution contains $M$ known gates. Let $\rho^{(t)}$ denote the final density matrix obtained by approximating the first $t$ known gates and keeping the remaining gates exact. Since the computations that define $\rho^{(t-1)}$ and $\rho^{(t)}$ differ only at location $t$, their states right before this gate are equal. The inequality above therefore gives a trace distance of at most $\epsilon_t$ after the gate. Since the subsequent operations cannot increase trace distance, it follows that
\[
 D(\rho^{(t)},\rho^{(t-1)})\le\epsilon_t.
\]
Using $\rho^{(0)}=\rho_{\rm out}$ and $\rho^{(M)}=\widetilde\rho_{\rm out}$, we obtain
\[
 D(\widetilde\rho_{\rm out},\rho_{\rm out})
 \le\sum_{t=1}^M D(\rho^{(t)},\rho^{(t-1)})
 \le\sum_{t=1}^M\epsilon_t.
\]
Applying a measurement to the final outputs gives the corresponding probability bound.
\end{proof}

To account for these errors, fix $0<\omega_{\rm gate}<1$ and apply Theorem~\ref{thm:main} with failure allowance $(1-\omega_{\rm gate})\delta$. If each known gate is approximated such that the operator-norm error is at most $\omega_{\rm gate}\delta/M$, the probability of failure increases by at most $\omega_{\rm gate}\delta$ and is therefore at most $\delta$. For fixed $\omega_{\rm gate}$, replacing $\delta$ by $(1-\omega_{\rm gate})\delta$ adds a constant to $\log(2/\delta)$, so the asymptotic query bounds are unchanged. The number of gates depends on the choice of synthesis method.

\section{Lower bound for expectation selection}
\label{app:expectation-lower}

We prove the $\Omega(\sqrt N/\eta)$ lower bound~\cite{WYLC} in the oracle model of Corollary~\ref{thm:expectations} by an adversary argument~\cite{Ambainis,HLS}.

\begin{proof}[Lower bound for Corollary~\ref{thm:expectations}]
Suppose first that $0<\eta\le1/8$. For each $j\in[N]$, define
\[
 W_j=\sum_{i=1}^N\ket i\bra i\otimes
 R_{\theta_{\mathbf1_{\{i=j\}}}},\qquad
 R_\theta=
 \begin{pmatrix}
 \cos\theta&-\sin\theta\\
 \sin\theta&\cos\theta
 \end{pmatrix},
\]
where $\theta_0=\pi/4$ and $\theta_1=\arcsin\sqrt{1/2-2\eta}$. With success projector $\ket1\bra1$, the objective value is $1/2-2\eta$ at label $j$ and $1/2$ at every other label. Since the difference is $2\eta$, an $\eta$-optimal output must be $j$.

Let $s_\eta=\|I-R_{\theta_1}^\dagger R_{\theta_0}\|$. The derivative of $\arcsin\sqrt{x}$ is at most $2/\sqrt3$ on $[1/4,1/2]$, and therefore
\[
 s_\eta
 =2\sin\bigl((\theta_0-\theta_1)/2\bigr)
 \le\theta_0-\theta_1
 \le\frac{4\eta}{\sqrt3}.
\]
We prove the lower bound with access to
\[
 \mathcal O_j
 =\sum_{\sigma\in\{-1,0,1\}}
 \ket\sigma\bra\sigma\otimes W_j^\sigma.
\]
The choices $\sigma=1,-1,0$ give a forward query, an inverse query, and the identity, respectively. Let $T$ be the maximum number of queries made by the algorithm. By postponing measurements until the end and completing shorter branches with identity queries ($\sigma=0$), the state of all registers is pure after each of the $T$ queries.

Write $\ket{\psi_j^t}$ for the state after $t$ queries on instance $j$, and let $\Gamma$ be the adjacency matrix of the complete graph on $[N]$. Define
\[
 \ket{\Phi_t}
 =N^{-1/2}\sum_j\ket j\ket{\psi_j^t},
 \qquad
 F_t=\bra{\Phi_t}(\Gamma\otimes I)\ket{\Phi_t}.
\]
The initial state is independent of $j$, so $F_0=N-1$. For $j'\ne j$, the algorithm returns $j$ with probability at least $2/3$ on instance $j$ and at most $1/3$ on instance $j'$. The trace distance between the final states is therefore at least $1/3$. Since these states are pure,
\[
 \bigl|\braket{\psi_j^T|\psi_{j'}^T}\bigr|
 \le\frac{2\sqrt2}{3},
 \qquad
 |F_T|\le\frac{2\sqrt2}{3}(N-1).
\]

To bound the change caused by a query, write
\[
 \mathcal Q=\sum_j\ket j\bra j\otimes\mathcal O_j,
 \qquad
 D=(\Gamma\otimes I)
   -\mathcal Q^\dagger(\Gamma\otimes I)\mathcal Q.
\]
The change in $F_t$ is at most $\|D\|$. For a forward query to label $i$, the work qubit is rotated by $R_{\theta_1}$ when $j=i$ and by $R_{\theta_0}$ otherwise. Ordering the instance indices with $i$ first gives the corresponding block of $D$ as
\[
 \begin{pmatrix}
 0&\mathbf1^{\mathsf T}\otimes C\\
 \mathbf1\otimes C^\dagger&0
 \end{pmatrix},
 \qquad
 C=I-R_{\theta_1}^\dagger R_{\theta_0},
\]
where $\mathbf1\in\mathbb R^{N-1}$ is the all-ones column vector. This block has norm $s_\eta\sqrt{N-1}$. Replacing $C$ by $C^\dagger$ gives the block for an inverse query. For $\sigma=0$, the query is the identity, so the corresponding block of $D$ is zero. Since $D$ is a direct sum over the queried labels and the choices of $\sigma$, it follows that
\[
 \|D\|=s_\eta\sqrt{N-1}.
\]
Operations independent of $j$ preserve $F_t$. Summing the changes over the $T$ queries therefore gives
\[
 \left(1-\frac{2\sqrt2}{3}\right)(N-1)
 \le |F_0-F_T|
 \le T s_\eta\sqrt{N-1}.
\]
Using $s_\eta\le4\eta/\sqrt3$, we obtain $T=\Omega(\sqrt N/\eta)$.

For $1/8<\eta<1$, take the exact Boolean value oracle with $f(j)=0$ and $f(i)=1$ for $i\ne j$. An $\eta$-optimal output must again be $j$. Each query, including a controlled or inverse query, can be implemented with a constant number of Boolean search queries. The search lower bound~\cite{BBBV} consequently gives
\[
 T=\Omega(\sqrt N)=\Omega(\sqrt N/\eta),
\]
which completes the proof.
\end{proof}

\section{Concrete constants and comparisons}\label{app:constants}

This appendix gives concrete constants for the parameters in the proposed algorithm, and compares it with a classical procedure based on medians and with the Chen--de Wolf construction in terms of query bounds. The parameters of each procedure are chosen so that the returned label $s$ and score $v$ satisfy
\[
 |v-f(s)|\le\eps,\qquad v\le\min_i f(i)+\eps
\]
with probability at least $15/16$.

\subsection{Parameter choices}\label{app:concrete-parameters}
In the analysis we consider the following exact rational constants:
\begin{align}
 (\gamma,\alpha,\beta)&=\left(\frac{17}{50},\frac7{20},\frac9{20}\right),&
 (\rho,h,m)&=\left(\frac7{10},\frac1{16},\frac1{18}\right),\nonumber\\
 (\vartheta,\theta)&=\left(\frac12,\frac9{26}\right),&
 (r_*,\Lambda)&=\left(\frac{14}{25},\frac1{200}\right),\nonumber\\
 (v_{\rm acc},\varepsilon_{\rm cap},\varepsilon_{\rm val},\sigma)
 &=\left(\frac{63}{64},\frac3{64},\frac1{64},\frac1{16}\right).
 \label{eq:concrete-constants}
\end{align}
These margins $(\gamma,\alpha,\beta)=(0.34,0.35,0.45)$ satisfy~\eqref{eq:margins} and give the signal gap $g=(\beta-\alpha)/(2-\alpha-\beta)=1/12$.

Use the polynomial family in Appendix~\ref{app:polynomial} with
\begin{equation}
 n_\ell=25,\qquad r_\ell=1+\lfloor\ell/2\rfloor,\qquad
 d_\ell=48r_\ell+1.
 \label{eq:concrete-kernels}
\end{equation}
A convenient error certificate is $E_{25,r}(1/12)\le u(r)$, where
\begin{equation}
 u(1)=\frac18,\qquad u(r)=\frac1{100}\,3^{-(r-2)}\quad(r\ge2).
 \label{eq:concrete-errors}
\end{equation}
Exact polynomial integration verifies $1\le r\le10$. For $r\ge11$, Lemma~\ref{lem:sign-polynomial} gives $E_{25,r}(1/12)\le(35/16)(25/12)^{-(2r-1)}\le u(r)$: the second inequality holds at $r=11$ and is preserved because $(12/25)^2<1/3$. Taking $\mathsf r_\ell=(1-u(r_\ell)/2)^4$ and $\mathsf b_\ell=(u(r_\ell)/2)^2$ therefore yields
\begin{equation}
 \prod_{\ell\ge0}\mathsf r_\ell
 \ge\left(\frac{15}{16}\right)^8\left(\frac{197}{200}\right)^4
 >\frac{14}{25},\qquad
 \sum_{j\ge1}\Lambda_j\le\frac1{247}<\frac1{200}.
 \label{eq:concrete-filter-certificate}
\end{equation}
Since $r_0=r_1=1$, we separate the first two factors in the retention product and apply $\prod(1-x_\ell)\ge1-\sum x_\ell$ to the remaining factors before taking the fourth power. The bound $\mathsf b_\ell\le1/256$ gives $\Lambda_j\le256^{-1}(9/256)^{j-1}$, from which the leakage estimate follows by summing over $j$. Moreover,
\[
 m=\frac1{18}<\frac7{125}=r_*(1-2\beta),
 \qquad
 \frac{h}{1-\rho}+\Lambda=\frac{16}{75}<\frac34,
\]
so the selection conditions are satisfied.

\subsection{Selection schedule}\label{app:concrete-schedule}
For any given $N$, use the fixed parameters above and the geometric schedule in Section~\ref{sec:geometric-schedule} with $p=1/N$ and $L_d=\log(1/\delta_d)$. In particular,
\begin{equation}
 K=1+\left\lceil\log_9\frac{9N}{8}\right\rceil,\qquad
 \delta_d=\frac3{64\,2^d},\qquad
 t_j=\left\lceil\frac{\log(1/\delta_d)}{\eta_d}\right\rceil,
 \quad d=K-j+1,
 \label{eq:concrete-schedule}
\end{equation}
where $\eta_d$ is given by~\eqref{eq:gain-schedule}. Let $T=\sum_jt_j$ and write $D(x\Vert y)=x\log(x/y)+(1-x)\log((1-x)/(1-y))$. Use the validation procedure of Lemma~\ref{lem:asymmetric-validation} with
\begin{equation}
 n_{\rm val}=\left\lceil\max\left\{
 \frac{\log(64T)}{D(\theta\Vert\gamma)},
 \frac{\log128}{D(\theta\Vert\alpha)}\right\}\right\rceil,
 \qquad k_{\rm val}=\lceil\theta n_{\rm val}\rceil.
 \label{eq:concrete-validation}
\end{equation}
Chernoff bounds give soundness at most $1/(64T)$ and completeness at least $63/64$. The total number of estimator queries, including validation, is bounded by
\[
Q_A=\sum_{j=1}^K t_jq_j^A+n_{\rm val}T,
\]
where $q_j^A$ is given in~\eqref{eq:trial-cost}. By Corollary~\ref{cor:geometric-seed}, this gives $Q_A=O(\sqrt N)$ for the chosen parameters.

For concrete choices outside this fixed-confidence comparison, one may use
\[
 (\omega_{\rm bad},\omega_{\rm none},\omega_{\rm rej})=(1/2,1/4,1/4)
\]
in Section~\ref{sec:confidence}, $c_{\rm gap}=1/4$ and $\sigma_E=1/8$ in Section~\ref{sec:applications}, and $\omega_{\rm gate}=1/2$ in Appendix~\ref{app:precision}.

\subsection{Comparison with baseline algorithms}\label{app:classical-baseline}
\label{app:cdw-constants}
\label{app:finite-comparison}
A natural classical comparison approach is based on computing the median of repeated estimates for each label. For each label $i$, let $v_i$ be the median of $k$ independent base-estimator outputs, where $k$ is odd. The procedure returns a label $s$ with $v_s=\min_i v_i$, together with the score $v=v_s$.

If $|v_i-f(i)|>\eps$, at least $(k+1)/2$ of the estimates used to form $v_i$ must be inaccurate. Since each estimate is inaccurate with probability at most $1/3$, it follows that $\Pr\{|v_i-f(i)|>\eps\}\le b_k$, where
\begin{equation}
 b_k=3^{-k}\sum_{r=(k+1)/2}^k\binom{k}{r}2^{k-r}.
 \label{eq:exact-median-tail}
\end{equation}
Choose the smallest odd $k$ satisfying $b_k\le1/(16N)$, and let $i_*$ be a label minimizing $f$. By the union bound, with probability at least $1-Nb_k\ge15/16$, all medians are accurate and therefore
\[
 |v-f(s)|\le\eps,\qquad
 v=v_s\le v_{i_*}\le f(i_*)+\eps.
\]
The procedure uses $Nk$ base-estimator queries.

Chen and de Wolf~\cite[arXiv v2, Theorem 2.4 and Appendix A]{CDW} give a bound of $M=1000\sqrt N\log_2(1/\delta_1)$ invocations of the amplified estimator and failure at most $\delta_1+M\sqrt{2b}$ when the amplified estimator has error at most $b$ per label. The logarithm is base two. We use their published prefactor and the simple choice $\delta_1=1/32$, giving $M=5000\sqrt N$. Take the smallest odd
\begin{equation}
 k\ge\frac{2\log(51200000000N)}{\log(9/8)}.
 \label{eq:published-cdw-median}
\end{equation}
The binomial Chernoff bound gives $b_k\le1/(51200000000N)$ and total failure at most $1/16$. A reversible median retaining its work uses $k$ base-estimator queries in either direction, so the budget is $Mk$. Neither these comparison parameters nor the proposed parameters in Appendix~\ref{app:concrete-parameters} are optimized.

Substituting the parameter values above into the query bounds gives the values in Table~\ref{tab:all-budgets}. The values are rounded up to 3 digits.
\begin{table}[!ht]
\centering\small
\caption{Worst-case base-estimator query bounds for success probability $15/16$ and selection error $2\eps$. Values are rounded upward to three significant figures.}
\label{tab:all-budgets}
\begin{tabular}{@{}lrrrrr@{}}
\toprule
Procedure & $N=2^{20}$ & $N=2^{40}$ & $N=2^{60}$ & $N=2^{80}$ & $N=2^{100}$\\
\midrule
Classical medians & $2.51\times10^{8}$ & $5.16\times10^{14}$ & $8.09\times10^{20}$ & $1.13\times10^{27}$ & $1.48\times10^{33}$\\
\addlinespace
CDW instantiation~\cite{CDW} & $3.36\times10^{9}$ & $4.68\times10^{12}$ & $6.04\times10^{15}$ & $7.49\times10^{18}$ & $9.00\times10^{21}$\\
\addlinespace
Proposed algorithm & $8.84\times10^{8}$ & $1.13\times10^{11}$ & $7.37\times10^{13}$ & $1.61\times10^{17}$ & $1.18\times10^{20}$\\
\bottomrule
\end{tabular}

\end{table}
The proposed algorithm gives the lowest query bound at each problem size in the table except for $N=2^{20}$, where the classical procedure gives the smallest bound.

For the proposed method, validation dominates the query budget for small problem sizes. Its share decreases from $84.55\%$ at $N=2^{20}$ to $0.00002794\%$ for $N=2^{100}$.

\subsection{Reducing the constants}

The constants in the query bounds could be reduced by using the finite amplification products in~\eqref{eq:finite-gain} and sharper retention and leakage estimates at each level. The filter accuracies and trial counts could then be chosen together with the probability margins and validation parameters to minimize
\[
 \sum_j t_jq_j^A+n_{\rm val}T.
\]
Since the relative cost of validation varies with $N$, it would be beneficial to optimize the parameters separately for each problem size $N$. Using exact binomial tails may reduce the number of estimates needed for validation compared with the Chernoff bounds used here.

It would be interesting to search numerically for minimax polynomials that satisfy the required approximation bounds with a lower degree. These polynomials must be odd and bounded in absolute value by one on $[-1,1]$. Another possibility is to combine the two cumulative-probability tests in Lemma~\ref{lem:clean} into one transformation. For the same analysis to be applicable, the resulting filter must satisfy the acceptance bounds and retain the scalar form of the initialized block and the workspace conditions of that lemma.

\bibliographystyle{quantum-notes}
\bibliography{references}
\end{document}